# A Hybrid APIM-CFGM Model for Longitudinal Non-Exchangeable Dyads: Demonstrating and Comparing Estimation Approaches Using Multilevel Modeling

Liu Liu

College of Education, University of Washington

## Abstract

Understanding change over time within dyads, such as mentor-mentee or therapist-client pairs, poses unique challenges, particularly in studies with small samples and distinguishable roles. This paper introduces a flexible hybrid longitudinal modeling that integrates features of the Actor-Partner Interdependence Model (APIM) and the Common Fate Growth Model (CFGM) to simultaneously estimate individual-level and shared dyad-level effects. Using a hypothetical peer-mentoring example (novices paired with experts), the model addresses three key issues: (1) how the interpretation of model parameters when role is dummy-coded versus effect-coded; (2) how model performance is affected by small sample sizes; and (3) how results differ between Maximum Likelihood (ML) and Bayesian estimation. Simulated data for 50 dyads across five time points are analyzed, with subsampling at 30 and 5 dyads. Models are estimated in a multilevel modeling framework using R (lme4 for ML and brms for Bayesian inference). Results show that dummy and effect coding reflect distinct interpretations: dummy coding expresses effects relative to a reference group, whereas effect coding centers parameters on the grand mean across roles. Although model fit remains unchanged, the choice of coding impacts how group-level effects and interactions are interpreted. Very small samples (e.g., 5 dyads) lead to unstable estimates, whereas Bayesian credible intervals more accurately reflect uncertainty in such cases. In larger samples, ML and Bayesian estimates converge. This APIM-CFGM hybrid offers a practical tool for researchers analyzing longitudinal dyadic data with distinguishable roles and smaller sample sizes.

**A Hybrid APIM-CFGM Model for Longitudinal Non-Exchangeable Dyads: Demonstrating and Comparing Estimation Approaches Using Multilevel Modeling**

### Introduction

The study of change over time is important for understanding developmental processes and responsiveness to intervention. For pairs of individuals, analyzing change requires specialized statistical techniques because dyadic data are simultaneously non-independent and represent the smallest possible cluster size (i.e., $n = 2$). An important analytic consideration is whether dyad members are exchangeable (or "indistinguishable") members who simply share a common context (e.g., two learners with the same tutor, two children from one family, or two students in the same classroom), or distinguishable, where each partner holds a defined role within the dyad (e.g., expert-novice, parent-child, or teacher-student). In many applied settings, distinguishable (non-exchangeable) dyads are assessed on common outcomes such as belonging, efficacy, or skill development, and understanding change in these contexts requires analytic strategies that can capture both individual-level effects and shared dyadic trajectories.

For non-exchangeable dyads in particular (also known as dyads with "distinguishable" members), two prominent models for cross-sectional data analysis have been used: the Actor-Partner Interdependence Model (APIM), wherein one member's predictor variable affects the other dyad member's response on an outcome variable (Kenny & Cook, 1999), and the Common Fate Model (CFM; Kenny & La Voie, 1985), which uses both members' score on a shared predictor to predict an outcome also shared by both members. In brief, APIM estimates individual-level (actor and partner) predictor effects on individual-level outcomes, whereas CFM estimates dyad-level predictor effects on dyad-level outcomes. These models offer complementary insights: APIM captures individual effects (how each person's characteristics influence outcomes), whereas CFM captures shared dyadic influences. Combining them yields a richer understanding of intra- and inter-personal processes (Wickham & Macia, 2019).

In practice, the APIM is implemented either as two separate univariate regressions (i.e., one for each dyad member), or as a covariance/structural path model to simultaneously capture relations for both dyad members (see Figure 1, Panel A). The CFM, however, requires estimation as a multivariate (structural equation) latent regression model with specialized residual error constraints to account for heterogeneity within dyads and dyad roles (see Figure 1, Panel B). These models have since evolved into cross-sectional hybrid APIM-CFM models (Figure 1,

Panels C and D; Wickham & Macia, 2019) and have also been extended to longitudinal applications using latent growth or autoregressive frameworks (e.g., Gistelinck & Loeys, 2019; Planalp et al., 2017; Iida et al., 2023). However, such models are typically implemented in costly software (e.g., Mplus, SAS) or complex syntax (e.g., using nlme in R), limiting accessibility. To address these challenges, the present study introduces a hybrid longitudinal APIM-CFGM model using a univariate two-level multilevel model (MLM) in the open-source R package lme4. This approach avoids complex residual specifications and supports use in small samples where traditional structural equation modeling (SEM)-based models may not be feasible, making it practical for applied researchers. The model simultaneously captures actor, partner, and common fate effects over time, while accommodating role-distinguishable dyads.

As part of this demonstration, we simulate data from a hypothetical peer mentoring program with expert and novice students and evaluate model performance under different conditions. Specifically, we address three methodological research questions: How does role coding (dummy vs. effect coding) influences parameter interpretability? How do model estimates behave as dyadic sample size decreases? And how does the estimation method (Maximum Likelihood vs. Bayesian) affect results? These questions are driven by practical concerns in applied dyadic research and are intended to guide researchers in making practical modeling decisions under real-world constraints.

**Motivation: Dyadic Data in Longitudinal Research**

In psychology, education, and the behavioral and social sciences, understanding how individuals change over time is often a core research aim. However, dyadic data introduce non-independence due to nesting, which can complicate longitudinal modeling. In some cases, such as reading interventions delivered in student pairs (e.g., Lü et al., 2023; Vadasy & Sanders, 2008), dyad members may be considered exchangeable if no role-based or demographic differences exist. In such cases, dyad membership can be treated as a nuisance factor, and standard three-level MLMs (time within person, persons within dyads) can be used. This allows estimation of growth parameters and predictors while accounting for within-person and within-dyad clustering. In other settings, non-exchangeable dyads involve distinct roles but only one member's behavior is modeled as the outcome, such as caregiver-patient pairs. For example, researchers might examine change in patients' mental health as a function of caregiver behaviors (Berridge et al., 2022a, 2022b, 2023), or vice versa. These are not truly dyadic models, as only one member's outcome is modeled; they reduce to two-level growth models with time nested in person. In contrast, the current study focuses on non-exchangeable dyads where both members are measured on the same outcome and differ in meaningful ways, such as role, experience, or demographics (Kenny et al., 2020). In peer mentoring, for instance, novice students may benefit differently than expert mentors with respect to college belonging or mental health (Plaskett et al., 2018). Such designs require more complex modeling because role cannot be ignored or treated as symmetrical; instead, it may be of interest to assess the degree to which dyad member role predicts change over time in each other's outcomes, and whether role interacts with individual- and dyad-level predictor effects on change.

**Traditional Dyadic Models for Cross-Sectional Designs**

*The Actor-Partner Interdependence Model*

The actor-partner interdependence model (APIM; Kenny, 1996) has been widely applied across psychological and social sciences, including studies on romantic relationship dynamics (e.g., Cook & Snyder, 2005) and attachment processes (e.g., Lozano et al., 2021). The APIM simultaneously estimates both intra-individual effects (i.e., actor effects) and inter-individual effects (i.e., partner effects), acknowledging that individuals within dyads influence each other's outcomes. Using our earlier college peer-mentoring example, APIM can address research questions such as: (1) Does a novice student's own perception of rapport with their partner predict their own feelings of belongingness? (actor effect); (2) Does the expert partner's perception of rapport predict the novice student's feelings of belongingness? (partner effect); and (3) Are both novice and expert perceptions of rapport uniquely predictive of each other's feelings of belongingness?

In practice, the APIM is a path model that regresses each dyad members' outcomes $Y_{Aj}$ and $Y_{Bj}$ on their own predictor (e.g., $X_{Aj}$) and their partner's predictor (e.g., $X_{Bj}$) simultaneously (Cook & Kenny, 2005; Wickham & Macia, 2019), as follows:

$$Y_{Aj} = \mu_A + a_{A(X)}X_{Aj} + p_{BA(X)}X_{Bj} + \varepsilon_{Aj} \tag{1}$$

$$Y_{Bj} = \mu_B + a_{B(X)}X_{Bj} + p_{AB(X)}X_{Aj} + \varepsilon_{Bj}, \tag{2}$$

where:

$j$: Dyad index, for $j$ = 1, 2, 3,…, $j$ dyads,

$A_j$: First member of the $j$-th dyad (e.g., expert),

$B_j$: Second member of the $j$-th dyad (e.g., novice),

$X_{A_j}, X_{B_j}$: Predictor variable for members A and B in dyad $j$, and

$\varepsilon_{A_j}, \varepsilon_{B_j}$: Residual errors, assumed to be normally distributed and possibly correlated.

Each individual's outcome ($Y_j$) is thus modeled as a function of their conditional mean ($\mu$), their own "actor" effect ($a_{(X)}$), their "partner's" effect ($p_{(X)}$), and correlated residual terms ($\varepsilon_j$). While each equation could be estimated separately, APIM is typically estimated simultaneously, often within a SEM framework to account for dyadic independence (see Figure 1, Panel A, for a path diagram).

### The Common Fate Model

The common fate model (CFM) also referred to as the "latent couple" or "shared fate" model, was introduced by Kenny (1996) and formalized for dyadic research as a way to model shared variance between partners. The basic CFM assumes that each dyad member's observed predictor is an indicator of a latent dyad-level predictor, and that each outcome is an indicator of a latent dyad-level outcome. This structure captures the dyad's shared "common fate." Essentially, the two members of the dyad contribute to common latent variables that influence outcomes for both (see Figure 1, Panel B, for a path diagram). Using the college peer-mentoring example again, CFM can answer research questions like: Does a dyad's shared level of rapport predict their shared outcome (belonging)? In doing so, CFM emphasizes dyad-level effects, modeling only the shared components of predictors and outcomes. Because CFM is a latent variable model, it typically requires SEM-specialized software or complex syntax, and it often assumes equal residual variances and covariances across partners to identify the model (Kenny & Ledermann, 2010). Unlike APIM, the CFM explicitly models correlated residuals to account for within-dyad and within-role dependencies (Cook, 1998; Ledermann & Kenny, 2012). Specifically, each dyad member's score serve as indicators of latent dyad-level variables, as follows (Wickham & Macia, 2019).

$$X_{Aj} = \nu_A + \lambda_A \xi_j + \delta_{Aj} \qquad (3)$$

$$X_{Bj} = \nu_B + \lambda_B \xi_j + \delta_{Bj} \qquad (4)$$

$$Y_{Aj} = \mu_A + \gamma_A \eta_j + \varepsilon_{Aj} \qquad (5)$$

$$Y_{Bj} = \mu_B + \gamma_B \eta_j + \varepsilon_{Bj} \qquad (6)$$

The structural component of the model at the dyad-level is:

$$\eta_j = \alpha + \psi\xi_j + \zeta_j \qquad . \qquad\qquad\qquad\qquad (7)$$

Here, $\lambda$ and $\gamma$ are factor loadings, $\delta$ and $\varepsilon$ are individual-specific residuals, $\xi_j$ and $\eta_j$ represent the latent predictor and outcome variables, respectively, for dyad $j$. The parameters $\nu_A$, $\nu_B$, $\mu_A$, $\mu_B$ are intercepts, which may differ by role (e.g., expert vs. novice), $\lambda_A$, $\lambda_B$, $\gamma_A$, $\gamma_B$ determine the extent to which each dyad member's observed score reflects the underlying latent dyad-level construct. The residuals $\delta_{Aj}$, $\delta_{Bj}$, $\varepsilon_{Aj}$, $\varepsilon_{Bj}$ capture individual-specific variability not explained by the latent factors and are typically allowed to correlate within dyads and/or within roles. In Equation (7), $\psi$ represents the association between the latent predictor $\xi_j$ and the latent outcome $\eta_j$, while $\zeta_j$ denotes the dyad-level residual in the outcome. For identification, the latent variables $\xi_j$ and $\eta_j$ are often standardized to have unit variance. Correlated residuals are used to account for non-independence both within each construct (between partners) and across constructs (within individuals).

While both APIM and CFM measure predictors and outcomes consistently across dyad members, the APIM uniquely focuses on individual-level relationships (intra- and inter-individual), whereas the CFM captures the mean relationship between a predictor and outcome at the dyad level, controlling for within-dyad dependencies. To accommodate both, recent extensions have combined APIM and CFM into hybrid models that capture both individual and dyadic influences. Cross-sectional hybrids (Wickham & Macia, 2019) include models where person-level predictors drive a latent dyad outcome (AP-CFM; Figure 1, Panel C) or where a latent dyad predictor drives individual outcomes (CF-APM; Figure 1, Panel D). These have been extended to longitudinal contexts using latent growth or autoregressive models (e.g., Gistelinck & Loeys, 2019, 2020; Iida et al., 2023; Ledermann & Macho, 2014; Planalp et al., 2017), but practical application remains limited by software and sample size constraints.

**Longitudinal Models for Dyadic Data**

The specification of longitudinal dyadic models largely depends on whether the research interest is (1) predicting change over time (e.g., growth models), (2) predicting length of time to reach an event (e.g., survival models), or (3) predicting the outcome mean across a series of measurements while controlling for prior measurements (e.g., autoregressive or cross-lagged panel models). Within dyadic methods research, the cross-sectional CFM has been adapted for the first type of research question (traditional growth models; Ledermann & Macho, 2014;

Planalp et al., 2017), while the cross-sectional APIM has been extended for autoregressive models (e.g., Gistelinck & Loeys, 2019, 2020). For example, Gistelinck and Loeys (2020) investigated how husbands' and wives' earlier intimacy perceptions affected each other's subsequent perceptions of intimacy. Because survival and autoregressive analyses are less focused on modeling growth parameters, and require additional assumptions (e.g., proportional hazards, stationarity), the present paper focuses on traditional growth models. Specifically, how dyad members' outcomes evolve over time and whether their trajectories diverge or converge. We propose and evaluate a hybrid longitudinal model that integrates APIM and CFM elements to simultaneously estimate individual-level and dyad-level effects across time.

### The Common Fate Growth Model

The Common Fate Growth Model (CFGM) focuses on estimating each type of dyad member's trajectory (i.e., intercept and slope) as well as predictors of those trajectories, in one joint model. The basic "default" model[1] that Planalp et al. (2017) describe does not include time-varying covariates, but still serves as a useful starting point, as follows.

For member $A$ (role A; e.g., expert) and member $B$ (role B; e.g., novice) in dyad $j$ at time $t$, the Level-1 (within-person) model is defined as:

$$Y_{Aj}(t) = \pi_{0Aj} + \pi_{1Aj}(Time_t) + r_{Aj}(t) \tag{8}$$

$$Y_{Bj}(t) = \pi_{0Bj} + \pi_{1Bj}(Time_t) + r_{Bj}(t) \tag{9}$$

where $Y_{Aj}(t)$ and $Y_{Bj}(t)$ are the observed outcome scores for members $A$ and $B$ of dyad $j$ at time $t$. The parameters $\pi_{0Aj}$ and $\pi_{0Bj}$ represent individual-specific intercepts (e.g., baseline levels), and $\pi_{1Aj}$ and $\pi_{1Bj}$ are individual-specific linear slopes. The residuals $r_{Aj}(t)$ and $r_{Bj}(t)$ capture within-person variability over time and are typically assumed to be normally distributed. Within a given time point, these residuals may be correlated between partners, and their variances may differ by role.

At Level-2 (between-dyads), these intercepts and slopes are modeled as:

---

[1] Although the optimal "dependent" CFGM with dependent errors, as featured in Planalp et al. (2017), can be estimated as a structural equation model (which assumes a fairly large dyadic sample size) or as a multilevel model with a constrained error covariance structure to incorporate heterogeneous role residual variances (though this structure is not available in the popular R lme4 package), the authors' "default" CFGM can be implemented in two alternative way. It can be specified as either a latent model as shown in Eq. 8 and 9 or as an analogous multilevel (univariate) model using R lme4 with nearly identical results. As such, we focus on the "default" CFGM as the basis for the APIM-CFM hybrid growth model, especially for analyzing small-sample dyadic longitudinal data.

$$\pi_{0Aj} = \gamma_{0A} + u_{0j} + u_{0Aj} \tag{10}$$

$$\pi_{0Bj} = \gamma_{0B} + u_{0j} + u_{0Bj} \tag{11}$$

$$\pi_{1Aj} = \gamma_{1A} + u_{1j} + u_{1Aj} \tag{12}$$

$$\pi_{1Bj} = \gamma_{1B} + u_{1j} + u_{1Bj} \tag{13}$$

where $\gamma_{0A}, \gamma_{0B}, \gamma_{1A}, \gamma_{1B}$ are the fixed effects (average intercepts and slopes for roles A and B across dyads). The terms $u_{0j}$ and $u_{1j}$ are dyad-level random effects shared by both members, representing random intercept and slope deviations for dyad $j$. The $u_{0Aj}, u_{0Bj}, u_{1Aj}, u_{1Bj}$ are role-specific random effects, allowing within-dyad individual differences between roles A and B in both intercept and slope. These random effects structure implies that the variance–covariance matrix of the latent growth parameters includes distinct variances for each member's intercept and slope, as well as covariances both within and across roles (e.g., correlations between partner intercepts and slopes).

The Level-2 model can be compactly expressed in vector form as:

$$\pi_j = \Gamma + U_j \tag{14}$$

where:

$$\pi_j = \begin{bmatrix} \pi_{0Aj} \\ \pi_{0Bj} \\ \pi_{1Aj} \\ \pi_{1Bj} \end{bmatrix}, \quad \Gamma = \begin{bmatrix} \gamma_{0A} \\ \gamma_{0B} \\ \gamma_{1A} \\ \gamma_{1B} \end{bmatrix}, \quad U_j = \begin{bmatrix} u_{0j} + u_{0Aj} \\ u_{0j} + u_{0Bj} \\ u_{1j} + u_{1Aj} \\ u_{1j} + u_{1Bj} \end{bmatrix}$$

The variance-covariance matrix, $U_j$ allows for distinct variance components for each member's intercept and slope, as well as covariance terms both within and across roles (e.g., intercept-slope correlations). This formulation (the "default" CFGM) does not include time-varying covariates.

When dyad members have distinct roles, a longitudinal growth model must allow for potentially different baseline levels and change rates for each role. The latent variable CFGM approach accomplishes this by estimating separate intercepts and slopes for each role as part of a multivariate SEM. Figure 2 illustrates a simplified CFGM path diagram for a distinguishable dyad measured at five time points. In such a model, each member (e.g., "expert" and "novice") has their own latent intercept and slope, but those latent factors are typically allowed to correlate across partners (capturing the dyad-level coupling). To achieve model identification and parsimony, CFGM implementations often constrain residual variances and covariances (e.g., assuming equal measurement error variance across partners and times, and no residual covariance between partners at the same time point). These constraints constitute a default error covariance structure in CFGM, ensuring that the shared latent growth factors capture most of the between-partner association, with minimal unexplained residual correlation.

While the latent CFGM is a powerful model for estimating dyadic trajectories, it requires specialized SEM software (e.g., Mplus) and may not converge reliably with small samples. To bridge these gaps, we propose a unified multilevel formulation of a hybrid APIM-CFGM model that retains the strengths of the latent CFGM while improving accessibility and small-sample performance. Implemented as a univariate two-level MLM using lme4 in R, this approach models repeated measurements nested within individuals, and individuals nested within dyads. Our approach uses a pairwise data structure (each dyad contributes two observations per time point, one per member) and includes role indicators as fixed effects. Dyad-level random intercepts and slopes capture the shared (common fate) components of growth, while role and role-by-time interactions capture differences between distinguishable members. This yields a multilevel growth model that mimics a CFGM via dyad-level random effects, while also allowing role-specific predictors and interactions like an APIM. In essence, the model estimates separate intercepts and slopes for each role (like APIM) but also includes random effects that are shared by partners (like CFGM), achieving a hybrid structure.

**The Univariate APIM-CFGM: A Hybrid Longitudinal Model for Non-Exchangeable Dyads**

To explicitly define our proposed model, we build on the basic CFGM by adding time-varying predictors that include each dyad members' own covariate (e.g., actor's rapport over time, partitioned into within- and between-person components) and their partners' covariate (e.g., partner rapport over time, similarly partitioned) to obtain APIM-CFGM hybrid results. Although the data structure technically has three levels (measurements, L1, within persons, L2, within dyads, L3), the non-exchangeability of dyad members via their dyadic "role" explains within-dyad variability when "role" is included as a fixed effect dummy- or effect-coded covariate. Specifically, we propose a two-step model in which Model 1 is the basic CFGM estimated as a 2-level univariate model, and Model 2 is the APIM-CFGM in which time-varying fixed effect covariates are incorporated, as follows. This hybrid approach conceptually merges the dyadic growth curve modeling and common fate modeling perspectives. Table A1 in Appendix summarizes the interpretation of model parameters under dummy coding, which was used in all reported models. For comparison, equivalent interpretations under effect coding are also included to highlight how coding choice affects parameter meaning.

Model 1 Basic CFGM (no covariates):

$$Y_{ijk} = \gamma_{000} + \gamma_{100} \cdot Time_{ijk} + \gamma_{200} \cdot Role_{jk} + \gamma_{300} \cdot (Time_{ijk} \times Role_{jk})$$
$$+ u_{00k} + u_{10k} \cdot Time_{ijk} + u_{20k} \cdot Role_{jk} + u_{30k} \cdot (Time_{ijk} \times Role_{jk}) + \varepsilon_{ijk} \quad (15)$$

Model 2 APIM-CFGM (with covariates):

$$Y_{ijk} = \gamma_{000} + \gamma_{100} \cdot Time_{ijk} + \gamma_{200} \cdot Role_{0jk} + \gamma_{300} \cdot (Time_{ijk} \times Role_{0jk})$$
$$+ \gamma_{400} \cdot X\_Actor\_WithinP_{ijk} + \gamma_{500} \cdot X\_Partner\_WithinP_{ijk}$$
$$+ \gamma_{600} \cdot (Time_{ijk} \times X\_Actor\_WithinP_{ijk}) + \gamma_{700} \cdot (Time_{ijk} \times X\_Partner\_WithinP_{ijk})$$
$$+ \gamma_{800} \cdot (Role_{0jk} \times X\_Actor\_WithinP_{ijk}) + \gamma_{900} \cdot (Role_{0jk} \times X\_Partner\_WithinP_{ijk})$$
$$+ \gamma_{1000} \cdot (Time_{ijk} \times Role_{0jk} \times X\_Actor\_WithinP_{ijk})$$
$$+ \gamma_{1100} \cdot (Time_{ijk} \times Role_{0jk} \times X\_Partner\_WithinP_{ijk})$$
$$+ \gamma_{001} \cdot X\_Actor\_Agg_{0jk} + \gamma_{002} \cdot X\_Partner\_Agg_{0jk}$$
$$+ \gamma_{101} \cdot (Time_{ijk} \times X\_Actor\_Agg_{0jk}) + \gamma_{102} \cdot (Time_{ijk} \times X\_Partner\_Agg_{0jk})$$
$$+ \gamma_{201} \cdot (Role_{0jk} \times X\_Actor\_Agg_{0jk}) + \gamma_{202} \cdot (Role_{0jk} \times X\_Partner\_Agg_{0jk})$$
$$+ \gamma_{301} \cdot (Time_{ijk} \times Role_{0jk} \times X\_Actor\_Agg_{0jk})$$
$$+ \gamma_{302} \cdot (Time_{ijk} \times Role_{0jk} \times X\_Partner\_Agg_{0jk})$$
$$+ U_{00k} + U_{10k} \cdot Time_{ijk} + U_{20k} \cdot Role_{0jk} + U_{30k} \cdot (Time_{ijk} \times Role_{0jk}) + \varepsilon_{i0k} \quad (16)$$

Assuming *Time* is coded such that the intercept is the first time point[2] and *Time* is coded in years, and that *Role* is dummy coded (e.g., novices = 1 and experts = 0), in both (15) and (16), the score at the $i^{th}$ timepoint within the $j^{th}$ person and $k^{th}$ dyad is a function of the sum of the: mean belongingness for experts at baseline ($\gamma_{000}$), the mean change per year on belongingness for experts ($\gamma_{100}$), the mean difference in baseline for the role coded 1 (novices) compared to the role coded 0 (experts) ($\gamma_{200}$), the difference in growth over time for role = 1 compared to role = 0 ($\gamma_{300}$), the deviation between the person's intercept and their dyad's intercept ($U_{00k}$), the deviation between the person's growth rate and their dyad's growth rate ($U_{30k}$), and the residual error of between the predicted and observed value at time $i$ for person $j$ in dyad $k$.

---

[2] In our demonstration, however, we coded time so that the intercept is the study midpoint rather than baseline.

Model 2 (Eq. 16) adds to Model 1 by incorporating two time-varying, centered covariates, *X_Actor_Within* (for the person's within-self variability) and *X_Partner_Within* (for the person's matched dyad member's within-person variability), and two aggregated, centered covariates, *X_Actor_Agg* and *X_Partner_Agg* (again, one for the person, and one for their partner). These respective four additional parameter slopes will estimate:

1. the predicted change in the outcome for the role coded 0 at time = 0 (baseline) for each 1-unit increase one's *own* covariate *value at time* = 0 (baseline), holding all else constant ($\gamma_{400}$) (within-person);

2. the predicted change in the outcome for the role coded 0 at time = 0 (baseline) for each 1-unit increase one's *own mean* covariate value across time points, holding all else constant ($\gamma_{001}$) (person-level aggregate);

3. the predicted change in the outcome for the role coded 0 at time = 0 (baseline) for each 1-unit increase in the one's *partner's* covariate *value at time* = 0 (baseline), holding all else constant ($\gamma_{500}$) (within-person); and

4. the predicted change in the outcome for the role coded 0 at baseline for each 1-unit increase in one's *partner's mean* covariate value across time points, holding all else constant ($\gamma_{002}$) (person-level aggregate).

The addition of *interactions between time and role with each of the two aforementioned time-varying (within-person) covariates* estimates the following quantities.

5. the predicted change in the mean growth rate for role = 0 for each 1-unit increase in one's own covariate relative to their covariate mean value, holding all else constant ($\gamma_{600}$);

6. the predicted change in the mean growth rate for role = 0 for each 1-unit increase in one's partner's covariate relative to their partner's covariate mean value, holding all else constant ($\gamma_{700}$);

7. the predicted difference between role = 1 compared to role = 0 in the relationship between the outcome and one's own covariate value at time = 0 (baseline), holding all else constant ($\gamma_{800}$);

8. the predicted difference between role = 1 compared to role = 0 in the relationship between the outcome and one's partner's covariate at time = 0 (baseline), holding all else constant ($\gamma_{900}$);

9. the predicted difference between role = 1 compared to role = 0 in the relationship between deviations in one's own covariate (relative to their mean value) and change over time in the outcome, holding all else constant ($\gamma_{1000}$); and

10. the predicted difference between role = 1 compared to role = 0 in the relationship between deviations in one's partner's covariate (relative to their mean value) and change over time in the outcome, holding all else constant ($\gamma_{1100}$).

Last, the incorporation of the *interactions among time and role with the two time-invariant covariates (i.e., person-level aggregates)* would estimate the following:

11. the predicted change in the mean growth rate for role = 0 per 1-unit increase in one's own mean covariate value, holding all else constant ($\gamma_{101}$);

12. the predicted change in the mean growth rate for role = 0 per 1-unit increase in one's partner's mean covariate value, holding all else constant ($\gamma_{102}$);

13. the predicted difference between role = 1 compared to role = 0 in the relationship between the outcome and one's own mean covariate value at the study midpoint, holding all else constant ($\gamma_{201}$);

14. the predicted difference between role = 1 compared to role = 0 in the relationship between the outcome and one's partner's mean covariate value at the study midpoint, holding all else constant ($\gamma_{202}$);

15. the predicted difference between role = 1 compared to role = 0 in the relationship between one's own mean covariate value and change over time in the outcome, holding all else constant ($\gamma_{301}$); and

16. the predicted difference between role = 1 compared to role = 0 in the relationship between one's partner's mean covariate value and change over time in the outcome, holding all else constant ($\gamma_{302}$).

**Trade-offs and Novelty of the Univariate Approach**

The APIM-CFGM hybrid introduced above is novel in that it achieves the goal of a dyadic latent growth model using MLM with random effects. This approach is both accessible, relying on open-source tools such as *lme4* and parsimonious, avoiding many of the assumptions and complexity associated with latent variable models. By modeling dyad-level random intercepts and slopes (e.g., $u_{00k}$, $u_{10k}$), the approach captures the "common fate" component of dyadic change over time without requiring complex residual covariance constraints, as required in the latent CFGM (Planalp et al., 2017). Role differences are handled through fixed effects (and optionally random effects), allowing researchers to model distinguishable role differences (e.g., mentor–mentee) without specifying separate latent factors for each role, as required in SEM.

A key trade-off of the MLM approach is its treatment of residual correlations. The model assumes independence of residuals $\varepsilon_{ijk}$ between partners at each time point, relying on dyad-level random effects to capture partner interdependence. If there are time-specific correlations beyond shared growth (e.g., moment-to-moment partner influence), MLMs cannot easily accommodate them unless custom residual structures or additional random effects are introduced. SEM, in contrast, allows residual correlations to be directly specified. Another trade-off involves distributional assumptions. MLM assumes normally distributed random effects and residuals, while SEM frameworks are more flexible in allowing, for example, heteroscedastic residuals across roles or time points. Nevertheless, for many applications where the primary interest lies in fixed effects and variance partitioning, the univariate MLM formulation provides a sufficiently rich yet streamlined alternative.

Nonetheless, this hybrid model is pragmatic for smaller samples. Latent variable models often require at least 100-200 dyads to reliably estimate variances and covariances (Kenny et al., 2006; Mehta & Neale, 2005; Planalp et al., 2017). In contrast, the multilevel approach can be estimated (with caution) in much smaller samples, especially using Bayesian estimation that can better quantify uncertainty with much smaller data. This hybrid model thus opens the door for researchers with modest dyadic samples (e.g., 20-30 dyads, or even fewer) to analyze longitudinal dyadic data with role differentiation, situations where previously one might have been forced to simplify the analysis or treat one member as just a "context" variable. To summarize, the APIM-CFGM hybrid model allows researchers to simultaneously examine: (a) differences between roles in baseline outcome levels and growth rates; (b) actor and partner effects of time-varying predictors on the outcome; (c) actor and partner effects of person-level predictors on overall outcome and on growth; and (d) whether any of these effects differ by role. All of this is achieved while properly accounting for the non-independence of dyad members through shared random effects, without the estimation burdens and sample size demands typically associated with fully latent variable frameworks.

**The APIM-CFGM in Context**

Going back to the peer mentoring example again: assuming belongingness (the outcome, $Y$) and rapport (the predictor, $X$) are measured five time points (such that rapport is treated as a time-varying covariate), and that Time was coded in years and centered at the midpoint of the study, the following research questions can be tested using the APIM-CFGM when role is coded with dummy coding (experts = 0, novices = 1).

1. For experts (role = 0) in the peer mentoring program, what is the mid-study (Time = 0) level of belongingness and how did it change over time?

2. How do novices (role = 1) in the peer mentoring program differ from experts in their mid-study level of belongingness, and in their growth rate in belongingness?

3. (a) Does a person's own perception of rapport at mid-study, and/or their partner's perception of rapport at mid-study, predict the expert's mid-study belongingness (i.e., treating rapport as a time-varying predictor at that time point)? (b) Do the dyad members' average rapport levels (across all time points) predict the expert's mid-study belongingness (treating rapport as a time-invariant predictor)? (c) Do these rapport variables (either at mid-study or on average) predict the expert's change over time in belongingness?

4. (a) Do effects of mid-study rapport on belongingness depend on dyad member role? In other words, do novices' mid-study rapport-belongingness relations differ from experts' mid-study rapport-belongingness relations? (b) Do effects of average rapport on mid-study belongingness depend on dyad member role? In other words, do novices' average rapport-belongingness relations differ from those of experts? (c) Last, do mid-study and/or average rapport effects on change over time in belongingness differ for novices and experts?

Importantly, the model's parameter estimates and their interpretations change if the dyad member role is effect-coded (i.e., expert = -1, novice = 1) instead of dummy-coded. Specifically, rather than positioning the role comparisons directly between novices and experts, all comparisons would be between the role coded 1 (novices) and the mean across roles. The intercept, for example, would be the average belongingness across both experts and novices (i.e., not the experts' mean), and similarly, the growth rate would be the average growth rate across both experts and novices. A dyadic, longitudinal approach like this allows us to examine both inter- and intra-personal dynamics within non-exchangeable dyads. For instance, prior research has shown significant wage disparities across race/ethnicity and gender in the early education workforce (Liu et al., 2023). This underscores the need for models that can capture how these intersectional factors interact over time within professional relationships. Moreover, prior studies, such as Lü et al. (2023) on educational settings, demonstrate the importance of understanding how different contextual factors, like home language backgrounds or race, influence educational frameworks. Together, these insights suggest that a longitudinal dyadic approach could help us to understand how such intersecting factors develop within professional relationships and impact both individual and joint outcomes in a variety of applied settings. In the forthcoming, we demonstrate the difference between dummy and effect coding dyad member role in the APIM-CFGM using our hypothetical example.

**Model Estimation Methods**

Maximum likelihood (ML) is the most widely used estimation method in multilevel and structural equation modeling. As a frequentist approach, ML estimates parameters by maximizing the probability that the observed data were generated by the model, assuming large-sample theory. However, ML's standard errors often underestimate uncertainty in small samples, inflating Type I error (Hox & McNeish, 2020). Restricted maximum likelihood (REML) can reduce bias in variance estimates but has limitations, such as less reliable convergence in complex models and inability to handle missing data as flexibly as ML (Baek et al., 2020). Another limitation of ML (and REML) with small samples is that core assumptions, such as normally distributed residuals and reliable asymptotic properties may not hold. Even when these assumptions are tenable, small samples lack the power to detect true effects. One common rule of thumb is that 100 cases are required for structural equation models using ML, which translates to 200 individuals in dyadic data (Kline, 2005). Others recommend a minimum of 200 cases (i.e., 400 observations in dyadic designs; Ledermann et al., 2022).

An increasingly popular alternative for small-sample modeling is Bayesian estimation (Gelman et al., 2013). Rather than relying on sampling distributions, Bayesian methods derive a posterior distribution by combining an empirical prior with the model likelihood. This posterior provides direct estimates of central tendency and uncertainty for each parameter, avoiding overconfidence often seen in small-sample ML estimates. Bayesian estimation requires the researcher to specify a prior distribution for each parameter. These may be uninformative (e.g., flat or uniform priors with wide ranges $\pm10{,}000$) or informative (e.g., normally distributed priors based on theory or prior data). Informative priors can improve precision in small samples but risk overpowering the data, especially when samples are sparse (Gelman, 2006). For this reason, some analysts prefer uninformative or weakly informative priors to avoid bias in estimation. Bayesian estimation uses Markov Chain Monte Carlo (MCMC) sampling to approximate posterior distributions by iteratively updating parameter estimates based on observed data and prior assumptions. MCMC algorithms typically include multiple chains, a burn-in period to discard early samples, and optional thinning to reduce autocorrelation. These routines are now streamlined in modern software packages such as Stan and brms. In the present study, we compare ML and Bayesian estimation across sample sizes to assess their relative performance for estimating a hybrid APIM-CFGM model in small dyadic datasets.

### Present Study

The current paper demonstrates the hybrid APIM-CFGM for analyzing change over time in outcomes using a hypothetical dataset for dyads with non-exchangeable members; the APIM-CFGM may be more practical than latent versions as well as MLMs with complex Level 1 error structures, especially for smaller samples of dyads. In addition, we also compare ML and Bayesian model estimation results. We specifically explore the following methodological research questions.

1. Interpretation under Different Role Coding: How are the hybrid model's parameters interpreted when the distinguishable roles are coded as dummy-coded (e.g., 0 = expert, 1 = novice) versus effect-coded (e.g., -1 = expert, +1 = novice)? What differences emerge in the estimates and their substantive meaning under these two coding schemes, and what are the implications for researchers in presenting and interpreting results?

2. Impact of Small Sample Size: How do the parameter estimates change when number of dyads becomes very small? In particular, what happens to fixed-effect estimates and variance components when we move from a reasonably sized sample (e.g., 50 dyads) to a much smaller sample (e.g., 5 dyads)?

3. Maximum Likelihood vs. Bayesian Estimation: Do Bayesian estimation techniques (with default priors) provide any advantages in terms of estimate stability or precision (interval width) compared to traditional ML estimation for this model, particularly as dyadic sample size becomes smaller? We compare the two estimation approaches to see if Bayesian credible intervals or point estimates differ meaningfully from ML results and whether Bayesian diagnostics reveal any estimation issues.

By investigating these questions, we aim to provide practical guidance on coding choices, sample size considerations, and estimation methods for applied researchers using the APIM-CFGM longitudinal model.

## Methods

### Data Simulation

While numerous dyadic datasets exist in psychology and education, few publicly available datasets contain longitudinal dyadic data where both members of a non-exchangeable dyad are measured repeatedly on the same outcome. This combination, including longitudinal structure, dyadic interdependence, and distinguishable roles is relatively rare in open-access repositories. Among those that do exist, many (1) do not include sufficient time points to estimate growth reliably, (2) use only one member's outcome (thus reducing the data to individual-level growth), or (3) are proprietary, incomplete, or require institutional access. Moreover, even when such datasets are accessible, they are often limited in their generalizability to the methodological questions posed in this study (e.g., role coding strategies, sample size effects, estimation comparisons).

To address these, we simulated longitudinal dyadic datasets data for this demonstration. The simulation was based on a hypothetical peer-mentoring program for first-generation college students (Duncheon, 2018, 2020). This context allowed to model intra- and inter-personal change in dyads over time, illustrating how the hybrid APIM-CFGM performs under conditions relevant to real-world intervention studies. We generated data for 50 dyads, each comprising one "novice" (first-year student) and one "expert" (continuing student), assessed over five time points (e.g., two academic years). One time-varying predictor (rapport) and one longitudinal outcome (college belongingness) were simulated as continuous, normally distributed variables. Both were standardized at baseline (mean = 0, $SD$ = 1), and role-specific linear trajectories were applied. Simulating in standardized units facilitates effect comparison and ensures results are not driven by scale artifacts. Using continuous normal variables aligns with multilevel linear model assumptions and simplifies interpretation. While real-world data may be skewed or ordinal, we intentionally focused on the ideal-normal case as a baseline; we return to generalizability in the Discussion. Data were initially generated in Mplus (Muthén & Muthén, 1998–2017), with additional between-dyad error added in R. The final dataset (in long format) is publicly available as a .csv file.

**Variables and Structure**

Each dyad was assigned a unique *dyadid*, with individuals identified by a *personid*. The variable *dyadrole* was coded in two ways: as a dummy-coded variable (*role_dum*), where 0 = expert and 1 = novice, and as an effect-coded variable (*role_eff*), where -1 = expert and 1 = novice. The outcome variable, *belong*, represents a hypothetical measure of college belongingness, and the predictor variable, *rapport*, represents a hypothetical individual-level perception of relational rapport. To separate within-person fluctuations from between-person differences, we computed person-mean-centered versions of *rapport*. Specifically, each person's rapport score at each time point was decomposed into: a within-person component (deviation from their own person mean), and a between-person component (their average rapport across time). We applied this procedure to both the individual (actor) and their partner. This yielded four key predictor variables: *rapport_actor_within* and *rapport_partner_within* (within-person deviations), and *rapport_actor_agg* and *rapport_partner_agg* (aggregate mean rapport). All aggregate variables were grand-mean-centered for use in regression models (so that intercepts represent average-case individuals). Because standard multilevel modeling software (e.g., lme4 in R) requires a univariate outcome structure, we reformatted the dataset into a pairwise (stacked) structure, as recommended by Bolger and Laurenceau (2013). In this structure, each row contains both the actor's and the partner's relevant covariates. For example, the value of *rapport_actor_within* from one row (e.g., 1.65 at the first time point for one partner) is assigned as *rapport_partner_within* in the row corresponding to their dyad partner. Similarly, an aggregate value such as *rapport_actor_agg* = 4.41 in one row appears as *rapport_partner_agg* in the partner's row. This cross-row mapping mirrors the structure used in multilevel mediation models (Bauer et al., 2006) and is illustrated in Figure 3.

Both the outcome and the predictor were assumed to be measured at five time points, hypothetically over a two-year academic period beginning in the early fall of year 1, then at the end of each of four semesters: fall in year 1, spring of year 1, fall of year 2, and spring of year 2 (at this point, the expert would theoretically graduate). Thus, the measurement-level predictor *time* was coded to reflect the distances from the middlemost point of the study (spring year 1), coded as -0.75, -0.5, 0, 0.5, and 1 years, respectively (Biesanz et al., 2004), which is a meaningful reference for interpretation (halfway through the mentoring program) and also reduces correlation between intercept and slope in the growth model.

**Small-sample Datasets**

　　　To explore model performance at smaller sample sizes, we drew random subsets of 30 and 5 dyads (using the *dplyr* package in R, with seed = 23). These subsamples reflect practical sample size challenges in applied research, especially when rarer populations are the focus (e.g., individuals with special needs or those from minoritized backgrounds). We selected $n$ =30 and $n$ = 5 dyads because the former size is often recommended for multilevel linear models for optimal standard error estimation, and the latter size is likely the smallest possible size one might use. Only one random subset for each condition was selected (rather than averaging over many simulations) to mimic the common scenario of a single study dataset available to researchers. As a result, some fluctuations observed at $n$ = 5 may reflect idiosyncratic of that specific sample. We acknowledge this limitation in the interpretation and emphasize that the results should be viewed as illustrative rather than definitive for all small-sample contexts.

　　　This paper fitted models described in Equations (15) and (16), corresponding to the basic CFGM and hybrid APIM-CFGM, using both ML and Bayesian estimation. Model 1 is a baseline two-level growth model (CFGM) with Time and Role predict the outcome variable, belongingness. Model 2 extends this to a full hybrid APIM-CFGM by incorporating the actor and partner rapport predictors (each decomposed into within- person and aggregate components) as well as their interactions with Time and Role. Both models include random intercepts and random slopes for Time at the dyad level to capture shared developmental trajectories, as well as random effects for Role and Role × Time (i.e., allowing role differences in intercepts and slopes to vary across dyads).

　　　ML estimation was implemented in R via the lme4 package (Bates et al., 2014), with Satterthwaite-adjusted degrees of freedom from lmerTest (Kuznetsova et al., 2017). For Bayesian estimation, we used the brms in R (Bürkner, 2017a, 2017b), which interfaces with Stan (Stan Development Team, 2020). The brms package was selected for its similarity to lme4 syntax and its capacity to flexibly estimate MLM. Each model was fit using 4 MCMC chains with 2,000 iterations per chain (1,000 warm-up), and no thinning beyond the default thin = 1. We use brms' default priors: (a) population-level regression coefficients (`class = "b"`, i.e., all slopes and interaction terms) were given an improper flat prior which is uniform over the real line ($\mathbb{R}$). This prior allows the data to fully determine the estimates without regularization; (b) the model intercept used the default Student-$t$(3, 0, 10) prior, which is weakly informative and centers the grand mean near zero while allowing wide dispersion; (c) standard deviations of group-level (dyad) random effects used the default half-Student-$t$(3, 0, 10) prior, providing gentle regularization for variance components; and (d) the random-effect correlation matrix followed the default LKJ ($\eta = 1$) prior, which is uniform over all correlation structures. We also used Stan's default initialization (uniform -2 to +2). Model convergence was assessed using multiple diagnostics. The Gelman-Rubin (R-hat) for all parameters was below 1.01, indicating chain convergence. Effective sample sizes (ESS) were high for $n = 30$ and $n = 50$ conditions, and acceptable for $n = 5$, though lower, as expected with small data. No divergent transitions or sampler warnings were observed, suggesting the posterior distribution was adequately explored in all conditions. Together, these diagnostics indicate that Bayesian models produced reliable posterior estimates both moderate and small sample sizes. While estimates converged with ML at larger $n$, the Bayesian approach offered more cautious uncertainty estimates in small samples, especially valuable when standard errors may be anti-conservative.

## Results

### Research Question 1: Model Interpretation for Dummy vs. Effect Coded "Role"

#### Dummy-Coded Role

Table 1 reports the fixed-effect estimates from both Model 1 (the basic CFGM) and Model 2 (full APIM-CFGM) under dummy coding (expert = 0, novice = 1), using both ML and Bayesian methods across the three sample sizes. For interpretive clarity, we focus on the largest sample ($n = 50$ dyads) and on ML estimates. Under dummy coding, model parameters have intuitive meanings with respect to the reference group (experts). For example, the intercept represents the estimated belongingness mean for experts at the midpoint of the study (Time = 0), and the Role coefficient indicates the difference in belongingness between novices and experts at time = 0. The Time × Role interaction captures how much faster (or slower) novices' belongingness change over time compared to experts. Actor and partner covariate effects are interpreted likewise with respect to experts. For instance, the actor-within-person effect reflects how fluctuations in an expert's rapport relate to their belongingness (at Time = 0), and the Role × Actor (within-person) interaction indicates whether that relationship differs for novices compared to experts. Interpretations for fixed and random effects under dummy coding are further elaborated in Tables 3 and 4, which detail how each model term should be read. Although Table 1 does not display random effects, their interpretations by coding scheme are provided in those supplementary tables. For reference, the residual (within-person) variance was estimated at $Var = 0.98$ in Model 1 and $Var = 0.95$ in Model 2 under ML estimation for $n = 50$ dyads.

### Effect-Coded Role

As described earlier, effect coding treats dyad member roles symmetrically by centering estimates on the grand mean (expert = -1, novice = +1). This approach does not alter the underlying model estimates or significance tests but changes how coefficients are interpreted (see Table 2). Specifically, intercepts represent the grand mean outcome (the average of experts and novices), and the Role coefficient reflects how far each role deviates from this grand mean (e.g., a positive coefficient indicates novices are above the mean, and experts below it by the same magnitude). Interaction terms reflect deviations from average effects rather than contrasts with a specific group. This distinction shifts the interpretive focus: effect coding answers questions like "How does each role differ from the overall mean?" whereas dummy coding answers "Compared with experts, how do novices differ?" Although the numerical estimates differ, key conclusions remain consistent across coding schemes. For example: novices showed significantly greater growth in belongingness than experts (positive Time × Role interaction), and an individual's own rapport (actor within-person covariate) positively predicted their belongingness, regardless of coding. No significant three-way interactions (e.g., Time × Role × Rapport) were found in the full sample. Again, Tables 3 and 4 include side-by-side interpretations for dummy and effect coding for all model parameters. Random effects under effect coding are also summarized in these tables, including variances centered on the grand mean rather than a specific group. For reference, the residual variance was estimated at $Var = 0.98$ in Model 1 and $Var = 0.95$ in Model 2 ($n = 50$ dyads for both models and ML estimation).

## Comparing Dummy vs. Effect Coding Results

Under dummy coding, we see that at $n = 50$, novices scored about 1.07 points higher in belongingness than experts at baseline (Role coefficient, not significant), and novices' growth rate was about 2.40 points per year higher than experts' (Time × Role, $p = .017$). The actor's within-person rapport effect is positive and significant ($0.36$, $p < .001$), indicating when an individual (expert or novice) felt higher rapport than usual, they tended to report higher belongingness at that time. None of the partner's rapport effects are significant in the full sample, suggesting one's partner's rapport (either momentary or on average) did not have a unique influence on one's belongingness when controlling for one's own rapport. No three-way interactions reached significance at $n = 50$, implying that the two-way interaction effects (e.g., Time × Role, or Role × Actor rapport) did not themselves differ over time or by role beyond the patterns already described.

In contrast, effect coding (see Table 2) would yield an intercept of about 2.01 (the grand mean belongingness at Time 0) and a Time coefficient of about 1.94 (the average slope across roles). The Role effect under effect coding (0.54) represents novices' deviation above the grand mean at Time 0 (which mirrors the dummy-coded estimate of 1.07 difference, just scaled by 1/2 since effect coding splits the difference between two roles). Importantly, the significance and magnitudes of key effects (e.g., the actor within-person rapport effect, the Time × Role interaction) remain the same; it is only their interpretation that changes. While the model's fit and conclusions remain consistent, the interpretation of coefficients shifts depending on the coding scheme.

Tables 3 and 4 offer detailed guidance on interpreting fixed and random effects under both dummy and effect coding. In general, dummy coding provides group-referenced interpretations and is especially useful when one role (e.g., expert) serves as a logical control or reference. Effect coding, by contrast, centers estimates on the grand mean and is more appropriate for symmetric designs or when both roles are of equal interest (Dalal & Zickar, 2012). In our study, dummy coding highlighted how novices differed from experts (e.g., faster growth in belongingness), while effect coding emphasized role deviations from the average dyad member. We recommend that applied researchers clearly report which coding scheme they use and, when possible, consider presenting results under both. In the remainder of our results, we report findings based on dummy-coded role for simplicity, noting where interpretation would differ under effect coding.

**Research Question 2: Model Performance in Smaller vs. Larger Samples**

To evaluate how the APIM-CFGM model performs with smaller samples, we estimated Model 2 (dummy-coded role) using three sample sizes: $n = 50$, 30, and 5 dyads. Table 1 summarizes the fixed effects under ML estimation for these samples, and Figure 4 (Panels A and B) plots the estimated coefficients and their standard errors by sample size. Results for the 30-dyad model closely mirrored those from the full 50-dyad sample: parameter estimates remained stable in direction and magnitude, and the pattern of statistical significance was preserved. For example, both the Time × Role interaction and the actor within-person effect remained significant at $n = 30$, with standard errors increasing modestly (by about 10-20%). These findings align with recommendations that roughly 30 clusters (dyads) often suffice for stable estimation of fixed effects and variance components in MLMs. In contrast, the 5-dyad model yielded unstable and potentially misleading results. Both point estimates and standard errors fluctuated substantially. For instance, the Time slope for experts increased from 0.80 (SE = 0.60) at $n = 50$ to 2.06 (SE = 0.96) at $n = 5$ with an inflated point estimate with high uncertainty. When combined with the Time × Role interaction, the implied slope for novices in the 5-dyad model was 4.3 points per year, compared to 3.2 at $n = 50$. This likely reflects idiosyncratic trajectories in a very small, unrepresentative sample.

Several other parameters showed similar instability. The main effect of role, while positive and nonsignificant at $n = 50$ (1.07, SE = 0.89), became even larger (1.68, SE = 1.37) at $n = 5$, though still non- significant due to the huge SE. The actor within-person effect declined from 0.36 (SE = 0.09) at $n = 50$ to 0.15 (SE = 0.25) at $n = 5$, becoming nonsignificant and less precise. Most notably, the partner within-person effect, near zero at $n = 50$, shifted to -0.76 (SE = 0.49) at $n = 5$, which if taken at face value would suggest a large negative partner influence. In reality, this is almost certainly overfitting to noise in the small sample. Random effect estimates were also unreliable at $n = 5$: one variance component collapsed to zero, suggesting the model could not distinguish true variance from noise, and residual variance estimates were erratic, further reflecting the model's difficulty extracting stable patterns from so few dyads.

In summary, the APIM-CFGM model produced stable and interpretable estimates with 30 dyads, suggesting that moderate samples may be sufficient for detecting medium-sized effects. However, estimates based on only 5 dyads were highly volatile. Under such conditions, MLMs may overfit, producing exaggerated, reversed, or spurious effects with inflated standard errors. We recommend simplifying the model, applying informative Bayesian priors, or using descriptive case-by-case approaches when analyzing very small dyadic samples. Any apparent effects in such tiny samples should be treated as exploratory and validated with additional data. Although this analysis used one random 5-dyad subsample, the observed instability is consistent with expected small-sample behavior: wide variability, unreliable estimates, and limited generalizability (Hox & McNeish, 2020).

**Research Question 3: Maximum Likelihood vs. Bayesian Estimation**

To evaluate differences between estimation methods, we compared Maximum Likelihood (ML) and Bayesian results for Model 2 using dummy-coded role (novice = 1, expert = 0) across three dyadic sample sizes. Table 1 includes both ML and Bayesian estimates side by side. At $n$ = 50 and $n$ = 30 dyads, Bayesian estimates closely matched ML estimates. Posterior means were identical to ML coefficients to the second decimal place, and 95% Bayesian credible intervals aligned with 95% ML confidence intervals. No parameter that was significant under ML become non-significant under Bayesian inference or vice versa. This consistency is expected under flat or weakly informative priors, where the likelihood dominates the posterior. At $n$ = 5, however, the results diverged. Posterior means remained similar to ML estimates (often slightly shrunk toward zero), but Bayesian credible intervals were substantially wider, reflecting greater uncertainty. As shown in Figure 5, Panel A, the point estimates under ML vs. Bayes still clustered along the diagonal line (indicating agreement in central values); but in Panel B, Bayesian posterior standard deviations exceeded ML standard errors for nearly all parameters. For instance, the actor within-person effect had an ML SE of 0.25 versus a Bayesian posterior SD of 0.40. Similarly, the Time slope for experts had an SE of 0.96 (ML) compared to 1.47 (Bayes), producing a broader interval. This divergence highlights a key distinction: ML relies on large-sample approximations that tend to underestimate uncertainty in small samples, whereas Bayesian estimation reflects the full joint uncertainty of all parameters. When data are very limited and priors are weak or flat, the posterior distribution becomes diffuse, appropriately signaling low information content.

All Bayesian models converged satisfactorily (R-hat = 1.00 for all parameters), though at $n$ = 5 several parameters had low effective sample sizes, consistent with very broad posteriors rather than MCMC failure. These results indicate that Bayesian estimation does not automatically improve precision in small samples unless (correct) informative priors are used. Instead, it provides a more transparent and conservative account of uncertainty. For example, in the 5-dyad model the ML fit suggested a partner within-person effect of -0.76 (SE = 0.49), which might (barely) be interpreted as a negative influence; the Bayesian 95% credible interval for that effect comfortably included zero, reinforcing that the apparent effect may reflect sampling noise. In sum, when sample sizes are sufficient (30-50 dyads), Bayesian and ML estimates converge. But with extreme data scarcity, Bayesian methods (with uninformative priors) yield wider, more cautious intervals that better reflect uncertainty, especially if priors are not strongly informative.

If researchers must analyze very small samples, leveraging informative priors grounded in theory or past data could stabilize estimates (Gelman, 2006); otherwise, the primary advantage of Bayesian analysis in this context is its honest portrayal of how little can be concluded.

## Discussion

This study introduced a practical hybrid dyadic growth model that integrates features of the Actor-Partner Interdependence Model (APIM) and the Common Fate Growth Model (CFGM) within a multilevel framework. The APIM-CFGM allows researchers to simultaneously assess intra-personal effects (how a person's own predictors influence their outcomes) and inter-personal effects (how a partner's predictors influence the individual's outcome), while also modeling role-based differences in baseline levels and change trajectories. This approach is well-suited to non-exchangeable dyads, such as mentor-mentee or teacher-student pairs, and is analytically feasible for small dyadic samples through MLMs and, when appropriate, Bayesian estimation.

### Research Applications

The hybrid APIM-CFGM model is not merely a methodological novelty, but a practical tool for dyadic research in education (e.g., teacher-student interventions), clinical psychology (e.g., therapist-client pairs), and organizational settings (e.g., mentor-protégé relationships). Whenever researchers have distinguishable dyads with both members measured on outcomes over time, this approach can be applied. For instance, in a teacher–student mentoring program, one could examine how each person's engagement (time-varying) predicts not only their own growth in outcomes but also their partner's, and how the mentor vs. student roles moderate these effects. The model's flexibility in R makes it accessible: users can fit it with lme4 (as we showed) or a Bayesian analog in brms, without requiring specialized SEM software. By providing annotated code and simulations, we hope to encourage researchers to explore hybrid dyadic models in their own work.

In our peer mentoring example, novices exhibited steeper increases in belongingness than experts. In real data, such a pattern could indicate that an intervention effectively supports newcomers, or conversely, flag role-based disparities in program impact. For instance, declining belongingness in one group might suggest mismatched pairs or unmet expectations. Even when overall mean change is minimal, large random-slope variance can reveal heterogeneity in individual change, highlighting a need to examine moderators such as motivation, personality, or contextual fit. Because the model incorporates time-varying covariates, researchers can ask: "Does feeling more rapport than usual at a given moment predict greater belongingness?" In our demonstration, an individual's momentary rapport predicted their concurrent belongingness, emphasizing the importance of relational quality at each occasion. While partner rapport had no effect in the simulated data, detecting such an influence in real-world contexts would point to a genuinely interpersonal mechanism. For example, a mentor's enthusiasm lifting the mentee's sense of inclusion beyond the mentee's own perceptions.

The model also supports role moderation. Analysts can ask: "Is rapport more consequential for novices than for experts?" In our case, the Role × Rapport interaction was nonsignificant, suggesting similar benefits across roles. However, a significant interaction in applied research could justify role-specific strategies, e.g., training mentors to foster rapport if it disproportionately benefits novices. Time interactions further enrich interpretation. For example, if rapport has a stronger effect early in the study than later, this could inform the timing of intervention delivery. Although such effects were null in our simulation, they may be present in real-world data. Taken together, the APIM-CFGM enables nuanced insight into dyadic change. Unlike a pre-post design, this model can isolate actor and partner effects, role differences, and heterogeneity in growth. By doing so, it offers a stronger basis for tailoring interventions. If novices respond more to rapport than experts, resources can be directed accordingly. If both partners follow a tightly coupled trajectory (a strong common fate component), joint support may be warranted. If one role lags, targeted boosts are appropriate.

**Methodological Implications**

This study offers several insights relevant to coding decisions, sample size, estimation strategy, and reporting practices in longitudinal dyadic models.

**Role Coding.** Choosing between dummy and effect coding for distinguishable dyads significantly impacts interpretation. While this is broadly recognized in regression contexts, it is especially consequential in dyadic research, where analysts must decide whether to treat one role as the reference or model roles symmetrically. Dummy coding provides direct, group-centered interpretations (e.g., "novices score 1.2 points higher than experts") and is intuitive when one group serves as a reference or is of primary interest (e.g., an experimental or at-risk group). In contrast, effect coding centers effects on the grand mean, placing both roles on equal footing and often simplifying the interpretation of complex interaction terms by orthogonalizing predictors.

The current paper echoes past recommendations (Dalal & Zickar, 2012) to report the coding scheme clearly, as different readers may find one or the other more intuitive. Tables 3 and 4 in our results serve as a template for how to translate between coding schemes for both fixed and random effects.

**Sample Size.** The small-sample analysis reinforces the importance of having an adequate number of dyads for reliable inference. While single-digit dyad counts might appear occasionally in literature (due to practical constraints), our findings add to a growing consensus (Maas & Hox, 2005; Hox & McNeish, 2020) that such extreme small $N$ can lead to wildly unstable estimates. If analyses must be conducted with very few dyads, results should be interpreted with extreme caution, and researchers should consider simpler models (e.g., reducing predictors, constraining effects), use of (correct) prior information, or even alternative analytical strategies (e.g., case studies or series of single-case analyses) until more data can be gathered.

**Estimation Strategy.** Comparing ML and Bayesian estimation revealed that Bayesian methods, on their own, do not inherently "solve" small-sample problems, unless analysts incorporate appropriate, informative priors. In our study, we used the brms package with default priors (diffuse for slopes and weakly informative for intercepts and variance components). While these defaults did not improve estimate precision in small samples, they did offer a meaningful advantage: by widening credible intervals, they helped avoid the overconfidence that often plagues ML estimates under data scarcity. In moderate-to-large samples, we found that ML and Bayesian point estimates were nearly identical, which should reassure users of lme4 that ML methods perform comparably to Bayesian approaches when prior information is not available. However, the Bayesian framework offers additional benefits, such as full posterior distributions, credible intervals that are often easier to interpret than frequentist confidence intervals, and convergence diagnostics (e.g., R-hat and effective sample size) that enhance both interpretation and model diagnostics. Taken together, our findings suggest that the greatest benefits of Bayesian methods in small samples emerge when researchers specify informative priors based on theory or empirical evidence. Without such priors, Bayesian estimation primarily serves to better reflect uncertainty, rather than reduce it. Nonetheless, the transparency and diagnostic tools available in Bayesian frameworks make them a complement to ML, particularly in challenging estimation scenarios.

**Fixed and Random Effects Interpretation.** The current paper shows how one can obtain a rich interpretation of model parameters by combining fixed-effect estimates with variance component estimates. We emphasize that even in applied reports, interpreting key random effects (variance components) is important. For example, quantifying how much dyad-to-dyad variability exists in outcomes and slopes, or how much of the outcome variance is attributable to role differences. In our example, we found that a substantial portion of variance (30%) was due to role differences in baseline outcomes (novices vs. experts), indicating the importance of accounting for role in the model. Similarly, around 36% of the variance in growth trajectories was due to role-based differences. These help contextualize the size of effects and could guide future studies in planning (e.g., how much variance one might expect a manipulation or intervention to explain, given baseline role differences).

**Model Complexity and Parsimony.** While our hybrid model captured a wide range of effects (all two-way and three-way interactions among role, time, and covariates), in practice a fully saturated model may be difficult to interpret or not necessary for every research question. Researchers should consider theoretical priorities and possibly trim or center certain interactions for parsimony. For example, if one is primarily interested in overall actor and partner effects and not their moderation by role over time, one might omit three-way interactions to simplify the model. Our approach is flexible and can be customized: one can include only those effects of interest (e.g., maybe only time-varying actor effects and a common fate random intercept). The code provided can be adapted accordingly.

## Limitations and Directions for Future Research

Several limitations of the present study suggest directions for future research. First, the model was demonstrated using a single simulated dataset with limited subsampling. While sufficient for illustration, the numerical results, particularly in the five-dyad condition are only exploratory. A comprehensive Monte Carlo simulation varying dyad count, effect size, ICC, and missingness would provide evidence-based guidelines (e.g., sample sizes needed to detect partner effects or stabilize variance estimates). Second, the Bayesian analyses relied on default brms priors, which include diffuse priors for regression terms and weakly informative priors for variance components. Although reasonable as defaults, more informative priors, which based on theory, meta-analysis, or expert input, may improve precision, especially in small samples. Future work should explore how prior specification affects estimation, particularly for random effects (e.g., Liu & Sanders, 2023; Liu, 2024). Third, the model does not account for higher-level clustering (e.g., classrooms, teams, therapy groups). Extending the APIM-CFGM to a three-level framework would enable tests of contextual moderation but would also raise sample size demands at each level. Fourth, only continuous, approximately normal outcomes and predictors were modeled. Many applied outcomes are categorical or count-based (e.g., diagnostic status, conflict episodes). The hybrid model can be generalized to a GLMM framework, though estimation becomes more complex in small samples. Bayesian methods with thoughtful priors may be especially helpful in such cases. Fifth, the full interaction structure may be difficult to interpret or overparameterized in practice. Researchers should consider theoretical priorities, centering strategies (e.g., midpoint centering of time), and model parsimony when selecting interaction terms. The full model offers a starting point, but trimmed versions may be preferable.

Despite these limitations, the hybrid APIM-CFGM expands the toolkit for longitudinal dyadic analysis. By combining actor-partner and common fate processes in a multilevel framework, the model enables researchers to examine how distinguishable dyad members influence one another over time. Annotated R code and practical modeling guidance (e.g., role coding, estimation, small samples) are provided to support use. As interest in interpersonal processes grows, tools like the APIM-CFGM will be essential for analyzing even modest dyadic longitudinal data.

**Data and Code Availability**

All datasets, annotated R scripts, model code, and supplementary materials have been deposited at the Open Science Framework (OSF). A private link has been provided to the journal during peer review, and the repository will be made publicly available upon acceptance of the manuscript.

**Note**

This is a preprint of a manuscript submitted for journal publication (July 2025). Posting this version on arXiv complies with the publisher's preprint policy. The content has not been peer reviewed or copyedited.

**Acknowledgements**

The author would like to thank Dr. Elizabeth Sanders for valuable feedback and guidance during the development of this work.

**Declarations**
**Funding**
This research received no specific grant from any funding agency in the public, commercial, or not-for-profit sectors.

**Conflicts of Interest**

The author declares no conflicts of interest.

**Ethics Approval**

Not applicable. This study used simulated data and did not involve human participants or identifiable information requiring ethics review.

**Consent to Participate**

Not applicable.

**Consent for Publication**

Not applicable.

**Table 1**

*Parameter Estimates for APIM-CFM using Dummy-Coded Role, by Dyad-Level Sample Size and Type of Model Estimation*

| Parameter | N = 50 | | | | N = 30 | | | | N = 5 | | | |
|---|---|---|---|---|---|---|---|---|---|---|---|---|
| | ML (Freq) | | Bayesian | | ML (Freq) | | Bayesian | | ML (Freq) | | Bayesian | |
| | Coeff | (SE) | Coeff | (SE) | Coeff | (SE) | Coeff | (SE) | Coeff | (SE) | Coeff | (SE) |
| *Model 1* | | | | | | | | | | | | |
| 1. Intercept (Expert Mean at Time = 0, Midpoint) | 1.48 | (0.59)* | 1.46 | (0.56)** | 1.89 | (0.74)* | 1.88 | (0.70)** | 1.61 | (0.76) | 1.48 | (1.13) |
| 2. Time (Expert Mean Change per Year) | 0.75 | (0.65)* | 0.73 | (0.63) | 0.79 | (0.85) | 0.78 | (0.81) | 2.06 | (0.96) | 2.02 | (1.47) |
| 3. Role (0 = Expert, 1 = Novice) | 1.07 | (0.89) | 1.07 | (0.85) | 0.56 | (1.12) | 0.52 | (1.07) | 1.68 | (1.37) | 1.57 | (1.47) |
| 4. Time * Role | 2.40 | (0.97) | 2.39 | (0.93)** | 1.93 | (1.20) | 1.89 | (1.14) | 2.32 | (1.82) | 2.22 | (2.06) |
| *Model 2* | | | | | | | | | | | | |
| 1. Intercept (Expert Mean at Time = 0, Midpoint) | 1.38 | (0.58)* | 1.35 | (0.59)* | 1.81 | (0.72)* | 1.80 | (0.71)* | 1.57 | (0.38)** | 1.47 | (1.05) |
| 2. Time (Expert Mean Change per Year) | 0.80 | (0.62) | 0.77 | (0.62) | 0.88 | (0.75) | 0.87 | (0.75) | 1.99 | (0.46)** | 2.02 | (1.22) |
| 3. Role (0 = Expert, 1 = Novice) | 1.35 | (0.85) | 1.35 | (0.84) | 0.73 | (1.06) | 0.69 | (1.05) | 1.75 | (0.35)** | 1.68 | (0.94) |
| 4. Actor Rapport, Within-Person (WP) | 0.36 | (0.08)*** | 0.35 | (0.08)*** | 0.36 | (0.10)*** | 0.36 | (0.11)** | 0.54 | (0.34) | 0.59 | (0.53) |
| 5. Actor Rapport, Aggregate (Agg) | 0.06 | (0.14) | 0.06 | (0.14) | 0.12 | (0.17) | 0.12 | (0.18) | -0.11 | (0.10) | -0.12 | (0.31) |
| 6. Partner Rapport, Within-Person (WP) | -0.10 | (0.09) | -0.10 | (0.09) | -0.05 | (0.12) | -0.05 | (0.13) | -0.96 | (0.46)* | -0.96 | (0.72) |
| 7. Partner Rapport, Aggregate (Agg) | -0.21 | (0.16) | -0.21 | (0.17) | -0.28 | (0.19) | -0.28 | (0.19) | -0.54 | (0.15)** | -0.50 | (0.49) |
| 8. Time * Role | 2.43 | (0.92)* | 2.44 | (0.91)** | 1.93 | (1.12) | 1.88 | (1.11) | 2.07 | (0.71)* | 1.98 | (1.70) |
| 9. Time * Actor Rapport WP | -0.02 | (0.04) | -0.02 | (0.04) | 0.00 | (0.05) | 0.01 | (0.05) | 0.17 | (0.10) | 0.16 | (0.14) |
| 10. Time * Actor Rapport Agg | -0.32 | (0.17) | -0.30 | (0.18) | -0.22 | (0.22) | -0.22 | (0.23) | -0.74 | (0.47) | -0.80 | (0.79) |
| 11. Time * Partner Rapport WP | 0.06 | (0.05) | 0.06 | (0.06) | 0.10 | (0.07) | 0.11 | (0.07) | 0.15 | (0.28) | 0.10 | (0.42) |
| 12. Time * Partner Rapport Agg | -0.24 | (0.19) | -0.25 | (0.20) | -0.34 | (0.23) | -0.34 | (0.23) | 0.20 | (0.50) | 0.23 | (0.87) |
| 13. Role * Actor Rapport WP | -0.18 | (0.12) | -0.19 | (0.13) | -0.26 | (0.16) | -0.27 | (0.17) | -0.55 | (0.60) | -0.65 | (0.89) |
| 14. Role * Actor Rapport Agg | 0.33 | (0.23) | 0.33 | (0.24) | 0.07 | (0.28) | 0.08 | (0.30) | 0.29 | (0.21) | 0.32 | (0.72) |
| 15. Role * Partner Rapport WP | 0.07 | (0.12) | 0.08 | (0.12) | 0.09 | (0.16) | 0.09 | (0.17) | 0.61 | (0.59) | 0.62 | (0.88) |
| 16. Role * Partner Rapport Agg | 0.15 | (0.22) | 0.15 | (0.24) | 0.20 | (0.28) | 0.20 | (0.30) | -0.37 | (0.20) | -0.41 | (0.68) |
| 17. Time * Role * Actor Rapport WP | 0.28 | (0.07) | 0.03 | (0.07) | 0.08 | (0.09) | 0.08 | (0.09) | 0.01 | (0.29) | 0.02 | (0.44) |
| 18. Time * Role * Actor Rapport Agg | 0.43 | (0.27) | 0.44 | (0.28) | 0.22 | (0.35) | 0.23 | (0.37) | 0.82 | (0.77) | 0.91 | (1.32) |
| 19. Time * Role * Partner Rapport WP | -0.05 | (0.07) | -0.05 | (0.28) | -0.06 | (0.09) | -0.07 | (0.09) | 0.06 | (0.29) | 0.12 | (0.45) |
| 20. Time * Role * Partner Rapport Agg | 0.28 | (0.27) | 0.27 | (0.28) | 0.23 | (0.34) | 0.23 | (0.37) | -0.63 | (0.74) | -0.68 | (1.25) |

*Note.* Role = dummy-coded (0 = Expert, 1 = Novice). Within-Person (WP) = within-person deviations into the centered version of the dyad members' rapport; Aggregate (Agg) = between-person means into the centered version of the dyad members' rapport; Time = -0.75, -0.5, 0, 0.5, 1 (Years).
* $p < .05$, ** $p < .01$, *** $p < .001$

**Table 2**

*Parameter Estimates for APIM-CFM using Effect-Coded Role, by Dyad-Level Sample Size and Type of Model Estimation*

| Parameter | $N = 50$ ML (Freq) Coeff | (SE) | $N = 50$ Bayesian Coeff | (SE) | $N = 30$ ML (Freq) Coeff | (SE) | $N = 30$ Bayesian Coeff | (SE) | $N = 5$ ML (Freq) Coeff | (SE) | $N = 5$ Bayesian Coeff | (SE) |
|---|---|---|---|---|---|---|---|---|---|---|---|---|
| *Model 1* | | | | | | | | | | | | |
| 1. Intercept (Mean at Time = 0, Midpoint) | 2.01 | (0.44)*** | 1.99 | (0.46)*** | 2.17 | (0.57)*** | 2.13 | (0.58)*** | 2.45 | (0.96) | 2.30 | (1.25) |
| 2. Time (Mean Change per Year) | 1.94 | (0.47)*** | 1.93 | (0.48)*** | 1.75 | (0.67)** | 1.72 | (0.67)* | 3.22 | (0.99)* | 3.12 | (1.43)* |
| 3. Role (-1 = Expert, 1 = Novice) | 0.54 | (0.44) | 0.52 | (0.46) | 0.28 | (0.56) | 0.27 | (0.56) | 0.84 | (0.68) | 0.83 | (1.00) |
| 4. Time * Role | 1.20 | (0.48)* | 1.18 | (0.50)* | 0.97 | (0.60) | 0.95 | (0.61) | 1.16 | (0.91) | 1.13 | (1.25) |
| *Model 2* | | | | | | | | | | | | |
| 1. Intercept (Mean at Time = 0, Midpoint) | 2.05 | (0.44)*** | 2.03 | (0.46)*** | 2.18 | (0.57)*** | 2.15 | (0.60)*** | 2.44 | (0.35)*** | 2.32 | (1.12) |
| 2. Time (Mean Change per Year) | 2.01 | (0.46)*** | 1.99 | (0.48)*** | 1.85 | (0.63)** | 1.82 | (0.66)** | 3.02 | (0.57)** | 2.99 | (1.72) |
| 3. Role (-1 = Expert, 1 = Novice) | 0.67 | (0.42) | 0.67 | (0.44) | 0.37 | (0.53) | 0.38 | (0.58) | 0.88 | (0.17)** | 0.88 | (0.66) |
| 4. Actor Rapport, Within-Person (WP) | 0.27 | (0.06)*** | 0.26 | (0.06)*** | 0.23 | (0.08)** | 0.22 | (0.09)** | 0.27 | (0.29) | 0.21 | (0.44) |
| 5. Actor Rapport, Aggregate (Agg) | 0.22 | (0.11)* | 0.22 | (0.11)* | 0.16 | (0.14) | 0.16 | (0.15) | 0.03 | (0.09) | 0.03 | (0.31) |
| 6. Partner Rapport, Within-Person | -0.07 | (0.06) | -0.06 | (0.06) | -0.01 | (0.08) | 0.00 | (0.08) | -0.66 | (0.29)* | -0.61 | (0.44) |
| 7. Partner Rapport, Aggregate (Agg) | -0.13 | (0.11) | -0.13 | (0.12) | -0.18 | (0.14) | -0.18 | (0.15) | -0.72 | (0.09)*** | -0.72 | (0.30)* |
| 8. Time * Role | 1.21 | (0.46)* | 1.22 | (0.48)* | 0.96 | (0.56) | 0.97 | (0.60) | 1.03 | (0.35)* | 1.00 | (1.04) |
| 9. Time * Actor Rapport WP | 0.00 | (0.03) | 0.00 | (0.04) | 0.05 | (0.04) | 0.05 | (0.05) | 0.17 | (0.15) | 0.16 | (0.21) |
| 10. Time * Actor Rapport Agg | -0.10 | (0.13) | -0.09 | (0.14) | -0.11 | (0.17) | -0.10 | (0.18) | -0.33 | (0.36) | -0.23 | (0.70) |
| 11. Time * Partner Rapport WP | 0.04 | (0.03) | 0.04 | (0.04) | 0.07 | (0.04) | 0.07 | (0.05) | 0.18 | (0.15) | 0.19 | (0.21) |
| 12. Time * Partner Rapport Agg | -0.10 | (0.13) | -0.11 | (0.14) | -0.22 | (0.17) | -0.23 | (0.19) | -0.11 | (0.35) | -0.16 | (0.69) |
| 13. Role * Actor Rapport WP | -0.09 | (0.06) | -0.09 | (0.06) | -0.13 | (0.08) | -0.13 | (0.09) | -0.27 | (0.30) | -0.30 | (0.44) |
| 14. Role * Actor Rapport Agg | 0.16 | (0.11) | 0.16 | (0.12) | 0.04 | (0.14) | 0.04 | (0.15) | 0.15 | (0.10) | 0.15 | (0.36) |
| 15. Role * Partner Rapport WP | 0.03 | (0.06) | 0.04 | (0.06) | 0.05 | (0.08) | 0.04 | (0.08) | 0.30 | (0.30) | 0.29 | (0.44) |
| 16. Role * Partner Rapport Agg | 0.07 | (0.11) | 0.07 | (0.12) | 0.10 | (0.14) | 0.10 | (0.15) | -0.18 | (0.10) | -0.18 | (0.36) |
| 17. Time * Role * Actor Rapport WP | 0.02 | (0.03) | 0.02 | (0.04) | 0.04 | (0.04) | 0.04 | (0.05) | 0.01 | (0.15) | 0.00 | (0.22) |
| 18. Time * Role * Actor Rapport Agg | 0.22 | (0.13) | 0.22 | (0.14) | 0.11 | (0.17) | 0.11 | (0.19) | 0.41 | (0.38) | 0.45 | (0.74) |
| 19. Time * Role * Partner Rapport WP | -0.03 | (0.03) | -0.03 | (0.03) | -0.03 | (0.04) | -0.03 | (0.04) | 0.03 | (0.15) | 0.02 | (0.21) |
| 20. Time * Role * Partner Rapport Agg | 0.14 | (0.13) | 0.14 | (0.14) | 0.12 | (0.17) | 0.12 | (0.18) | -0.31 | (0.37) | -0.33 | (0.72) |

*Note.* Role = effect-coded (-1 = Expert, 1 = Novice). Within-Person (WP) = within-person deviations into the centered version of the dyad members' rapport; Aggregate (Agg) = between-person means into the centered version of the dyad members' rapport; Time = -0.75, -0.5, 0, 0.5, 1 (Years).
* $p < .05$, ** $p < .01$, *** $p < .001$.

**Table 3**

*Interpretation of Fixed and Random Effects for the Basic CFGM Model (Model 1) Under Dummy Coding and Effect Coding*

| | Fixed Effects | Dummy Coding Interpretation | Effect Coding Interpretation |
|---|---|---|---|
| 1 | Intercept | For experts (the reference role, coded 0) in the peer mentoring program, the estimated mean level of belongingness at the midpoint of the study (Time = 0) was 1.48, holding all else constant. This effect was statistically significant ($p$ = .015). | In the peer mentoring program, the estimated mean level of belongingness across both dyad member roles at the midpoint of the study (Time = 0) was 2.01, holding all else constant. This effect was statistically significant ($p$ < .001). |
| 2 | Time | For experts (the reference role, coded 0), belongingness was predicted to increase by 0.75 points per year, holding all else constant. This effect was not statistically significant ($p$ = .258). | On average across both dyad members, belongingness was predicted to increase by 1.94 points per year, holding all else constant. This effect was statistically significant ($p$ < .001). |
| 3 | Role | At the midpoint of the study (Time = 0), novices (role = 1) were predicted to report 1.07 points higher belongingness than experts (role = 0), holding all else constant. This difference was not statistically significant ($p$ = .231). | At the midpoint of the study (Time = 0), novices (role = +1) were predicted to report 0.54 points higher belongingness than the grand mean across both roles, holding all else constant. This effect was not statistically significant ($p$ = .231). |
| 4 | Time*Role | Novices (role = 1) were predicted to experience a 2.40-point greater increase in belongingness per year than experts (role = 0), whose growth rate was 0.75 points per year, holding all else constant. This interaction was statistically significant ($p$ = .017). In other words, novices' predicted growth rate was 3.15 points per year (0.75 + 2.40), compared to 0.75 for experts. | Novices (role = +1) were predicted to experience 1.20 points more growth per year in belongingness than the average growth rate across both roles (M = 1.94), holding all else constant. This interaction was statistically significant ($p$ = .017). In other words, novices' predicted growth rate was 3.14 points per year (1.94 + 1.20), while experts' growth rate was 0.74 points per year (1.94 - 1.20). |

| | Random Effects (Variances) | Dummy Coding Interpretation | Effect Coding Interpretation |
|---|---|---|---|
| 1 | Intercept variance | Dyad-level variability in belongingness at the midpoint of the study (Time = 0) for experts (role = 0) was estimated at Var = 16.91, indicating that approximately 14% of the variance in experts' mid-study belongingness was attributable to dyad membership. | Dyad-level variability in belongingness at the midpoint of the study (Time = 0), averaged across both roles, was estimated at Var = 9.37, indicating that approximately 22% of the variance in mid-study belongingness was attributable to dyad membership. |
| 2 | Slope (Time) variance | The variability in yearly growth rates for experts (role = 0) was estimated at Var = 20.81, indicating that approximately 17% of the variance in belongingness was attributable to differences in experts' individual growth trajectories over time. | The variability in yearly growth rates, averaged across both roles, was estimated at Var = 10.62, indicating that approximately 25% of the variance in belongingness was attributable to individual differences in change over time. |
| 3 | Role variance | The variance in belongingness attributable to role differences within dyads (novices vs. experts) at the midpoint of the study (Time = 0) was estimated at Var = 38.86, indicating that approximately 31% of the total variance in mid-study belongingness was explained by systematic differences between roles. | The variance in belongingness attributable to role-related deviations from the grand mean at the midpoint of the study was estimated at Var = 9.71, indicating that approximately 23% of the total variance in belongingness was explained by role differences within dyads. |
| 4 | Time*Role variance | The variance in growth rate differences between novices (role = 1) and experts (role = 0) was estimated at Var = 45.98, indicating that approximately 37% of the variance in belongingness growth trajectories was attributable to role-based differences in change over time. | The variance in growth rate differences for novices (role = +1) relative to the average growth rate across roles was estimated at Var = 11.50, suggesting that approximately 27% of the variance in belongingness growth trajectories was attributable to role-based deviations from the average slope. |

*Note.* All fixed effect interpretations are evaluated at the midpoint of the study (*Time* = 0). Dummy coding was used in primary estimation (expert = 0, novice = 1); effect coding comparisons are included for interpretive contrast (expert = +1, novice = -1). Random effect variances reflect between-dyad or role-based variability as specified in the model.

**Table 4**

*Interpretation of Fixed and Random Effects for Model 2 (Hybrid APIM-CFGM) Under Dummy and Effect Coding*

| | Fixed Effects | Dummy Coding Interpretation | Effect Coding Interpretation |
|---|---|---|---|
| 1 | Intercept | For experts (role = 0) in the peer mentoring program, the estimated mean level of belongingness at the midpoint of the study (Time = 0) was at 1.38 points, holding all else constant. This effect was statistically significant ($p = .022$). | In the peer mentoring program, the estimated mean level of belongingness across dyad members at the midpoint of the study (Time = 0) was 2.05 points, holding all else constant. This effect was statistically significant ($p < .001$). |
| 2 | Time | For experts (role = 0), feelings of belongingness were predicted to increase by 0.80 points per year, holding all else constant. This growth rate was not statistically significant ($p = .205$). | On average across dyad members, the predicted growth rate in belongingness was 2.01 points per year, holding all else constant. This effect was statistically significant ($p < .001$). |
| 3 | Role | Novices (role = 1) in the peer mentoring program were estimated to score 1.35 points higher in belongingness at the midpoint of the study (Time = 0) than experts (role = 0), all else held constant. This difference was not statistically significant ($p = .119$). | Novices (role = +1) were estimated to score 0.67 points higher in belongingness at the midpoint of the study (Time = 0) compared to the grand mean across both roles, holding all else constant. This deviation was not statistically significant ($p = .119$). |
| 4 | Actor Rapport, Within-Person Centered (WP) | For experts (the reference role, coded 0), a 1-point increase in one's own rapport rating at the study midpoint (relative to their own average rapport across time) was associated with a 0.36-point increase in belongingness, holding all else constant. This effect was statistically significant ($p < .001$). In other words, when experts rated their rapport higher than usual, their mid-study belongingness tended to be higher as well. | On average across both roles, a 1-point increase in one's own rapport rating at the study midpoint (relative to their own average rapport across time) was associated with a 0.27-point increase in belongingness, holding all else constant. This effect was statistically significant ($p < .001$). That is, higher-than-usual rapport was generally predictive of higher belongingness at the midpoint of the study. |
| 5 | Actor Rapport, Aggregated and Centered (Agg) | For experts (the reference role, coded 0), a 1-point increase in one's mean rapport level (averaged across all time points) was associated with a 0.06-point increase in mid-study belongingness, holding | On average across both roles, a 1-point increase in one's mean rapport level (averaged across all time points) was associated with a 0.22-point increase in mid-study |

| | Fixed Effects | Dummy Coding Interpretation | Effect Coding Interpretation |
|---|---|---|---|
| | | all else constant. This effect was not statistically significant ($p$ = .669). That is, for experts, differences in typical rapport levels were not predictive of belongingness at the midpoint of the study. | belongingness, holding all else constant. This effect was statistically significant ($p$ = .050). In other words, individuals with higher typical rapport levels tended to report greater belongingness at the study midpoint. |
| 6 | Partner Rapport, Within-Person Centered (WP) | For experts (the reference role, coded 0), a 1-point increase in one's partner's rapport at the study midpoint (relative to the partner's typical rapport) was associated with a 0.10-point decrease in belongingness, holding all else constant. This effect was not statistically significant ($p$ = .255). In other words, momentary increases in partner rapport were not reliably associated with changes in experts' mid-study belongingness. | On average across both roles, a 1-point increase in one's partner's rapport at the study midpoint (relative to their typical rapport) was associated with a 0.07-point decrease in belongingness, holding all else constant. This effect was not statistically significant ($p$ = .248). That is, across both roles, there was no meaningful relationship between partner rapport at mid-study and mid-study belongingness. |
| 7 | Partner Rapport, Aggregated and Centered (Agg) | For experts (the reference role, coded 0), a 1-point increase in one's partner's mean rapport rating (averaged across time points) was associated with a 0.21-point decrease in belongingness at the study midpoint, holding all else constant. This effect was not statistically significant ($p$ = .210). That is, differences in a partner's typical (trait-level) rapport were not reliably associated with experts' mid-study belongingness. | On average across both roles, a 1-point increase in one's partner's mean rapport rating (averaged across time points) was associated with a 0.13-point decrease in mid-study belongingness, holding all else constant. This effect was not statistically significant ($p$ = .232). In other words, across both roles, partner typical rapport levels were not predictive of differences in mid-study belongingness. |
| 8 | Time*Role | Novices (role = 1) were predicted to experience 2.43 points more growth in belongingness per year than experts (role = 0), holding all else constant. This difference in slope was statistically significant ($p$ = .011). Specifically, novices' predicted growth rate was 3.23 points per year (0.80 + 2.43), compared to 0.80 points per year for experts. | Novices (role = +1) were predicted to have 1.21 points more growth in belongingness per year relative to the average growth rate across both roles, holding all else constant ($p$ = .011). This implies a predicted growth rate of 3.22 points per year for novices (2.01 + 1.21) and 0.80 points per year for experts (2.01 - 1.21). |

| | Fixed Effects | Dummy Coding Interpretation | Effect Coding Interpretation |
|---|---|---|---|
| 9 | Time*Actor Rapport (WP) | For experts (the reference role, coded 0), a 1-point increase in one's own rapport rating (relative to their mean rapport across time) was associated with a 0.02-point decrease per year in the predicted growth of belongingness, holding all else constant. This interaction effect was not statistically significant ($p = .727$). In other words, deviations from one's typical rapport were not predictive of changes over time in experts' belongingness. | On average across both roles, a 1-point increase in one's own rapport rating (relative to their mean rapport across time) was not associated with any change in the predicted growth rate of belongingness, holding all else constant. This effect was not statistically significant ($p = .968$). That is, momentary changes in one's own rapport were not predictive of differences in growth rate in belongingness across roles. |
| 10 | Time*Actor Rapport (Agg) | For experts (the reference role, coded 0), a 1-point increase in one's mean rapport rating (averaged across time points) was associated with a 0.32-point decrease per year in the predicted growth rate of belongingness, holding all else constant. This effect was not statistically significant ($p = .073$). In other words, a person's average rapport level was not reliably predictive of changes over time in experts' belongingness. | On average across both roles, a 1-point increase in one's mean rapport rating (averaged across time points) was associated with a 0.10-point decrease per year in the predicted growth rate of belongingness, holding all else constant. This effect was not statistically significant ($p = .463$). That is, trait-level rapport was not meaningfully related to changes in belongingness over time, on average across roles. |
| 11 | Time*Partner Rapport (WP) | For experts (the reference role, coded 0), a 1-point increase in one's partner's rapport rating (relative to their own mean rapport across time) was associated with a 0.06-point increase per year in the predicted growth of belongingness, holding all else constant. This effect was not statistically significant ($p = .244$). In other words, momentary increases in a partner's rapport were not predictive of changes in experts' belongingness over time. | On average across both roles, a 1-point increase in one's partner's rapport rating (relative to their own mean rapport across time) was associated with a 0.04-point increase per year in the predicted growth rate of belongingness, holding all else constant. This effect was not statistically significant ($p = .283$). That is, within-person fluctuations in partner rapport were not meaningfully related to average change in belongingness over time. |
| 12 | Time*Partner Rapport (Agg) | For experts (the reference role, coded 0), a 1-point increase in one's partner's mean rapport rating (averaged across time points) was associated with a | On average across both roles, a 1-point increase in one's partner's mean rapport rating (averaged across time points) was associated with a 0.10-point decrease per year |

|  | Fixed Effects | Dummy Coding Interpretation | Effect Coding Interpretation |
|---|---|---|---|
|  |  | 0.24-point decrease per year in the predicted growth rate of belongingness, holding all else constant. This effect was not statistically significant ($p$ = .207). In other words, a partner's typical rapport level was not reliably predictive of changes in experts' belongingness over time. | in the predicted growth rate of belongingness, holding all else constant. This effect was not statistically significant ($p$ = .434). That is, a partner's trait-level rapport was not meaningfully related to average change in belongingness over time. |
| 13 | Role*Actor Rapport (WP) | The relationship between novices' (role = 1) mid-study rapport and their mid-study feelings of belongingness was estimated to be 0.18 points lower than that observed for experts (role = 0), whose actor effect was 0.36 points, holding all else constant. This difference was not statistically significant ($p$ = .131). In other words, novices and experts did not significantly differ in the strength of the association between their own rapport and belongingness at the study midpoint. | The relationship between novices' mid-study rapport and belongingness was estimated to be 0.09 points lower than the average actor effect across both roles, holding all else constant. This difference was not statistically significant ($p$ = .131). That is, the association between one's own rapport and belongingness at mid-study did not significantly vary by dyad role. |
| 14 | Role*Actor Rapport (Agg) | The relationship between novices' (role = 1) mean rapport (averaged across time points) and mid-study belongingness was estimated to be 0.33 points higher than that observed for experts (role = 0), whose actor effect was 0.06 points, holding all else constant. This difference was not statistically significant ($p$ = .152). In other words, novices and experts did not significantly differ in how their typical rapport levels related to mid-study belongingness. | The relationship between novices' mean rapport and mid-study belongingness was estimated to be 0.16 points higher than the average actor effect across both roles, holding all else constant. This difference was not statistically significant ($p$ = .152). That is, the association between one's typical rapport and belongingness at mid-study did not significantly vary by dyad role. |
| 15 | Role*Partner Rapport (WP) | The relationship between novices' (role = 1) partners' mid-study rapport and their mid-study feelings of belongingness was estimated to be 0.07 points higher than that observed for experts' partners (role = 0), whose partner effect was 0.10 | The relationship between novices' partners' mid-study rapport and their belongingness was estimated to be 0.03 points higher than the average partner effect across both roles, holding all else constant. This difference was not statistically significant ($p$ = .579). |

| | Fixed Effects | Dummy Coding Interpretation | Effect Coding Interpretation |
|---|---|---|---|
| | | points, holding all else constant. This difference was not statistically significant ($p$ = .579).<br>In other words, the strength of the association between partner rapport and belongingness did not significantly differ between novices and experts. | That is, the association between a partner's rapport and one's own mid-study belongingness did not significantly vary by dyad role. |
| 16 | Role*Partner Rapport (Agg) | For novices (role = 1), the relationship between their partner's mean rapport (averaged across time points) and their own mid-study belongingness was estimated to be 0.15 points higher than the relationship observed for experts' partners (role = 0), whose partner effect was -0.21 points, holding all else constant. This difference was not statistically significant ($p$ = .509).<br>In other words, the association between a partner's typical rapport and mid-study belongingness did not significantly differ between novices and experts. | For novices (role = +1), the relationship between their partner's mean rapport and mid-study belongingness was estimated to be 0.07 points higher than the average partner effect across both roles, holding all else constant. This difference was not statistically significant ($p$ = .509).<br>That is, the strength of the association between partner's typical rapport and one's own belongingness at mid-study did not vary significantly by role. |
| 17 | Time*Role*Actor Rapport (WP) | The difference between novices' and experts' belongingness growth rates was predicted to increase by 0.03 points per year for each 1-point increase in one's own rapport rating at a given time point (relative to their mean rapport across time), holding all else constant. This three-way interaction was not statistically significant ($p$ = .634).<br>In other words, fluctuations in one's own rapport were not predictive of differential growth in belongingness between novices and experts. | Relative to the average growth rate across both roles, novices' belongingness growth rate was predicted to increase by 0.02 points per year for each 1-point increase in their time-specific rapport (relative to their mean), holding all else constant. This effect was not statistically significant ($p$ = .634).<br>That is, momentary increases in one's own rapport did not significantly affect the role-based difference in growth rate. |
| 18 | Time*Role*Actor Rapport (Agg) | The difference between novices' and experts' belongingness growth rates was predicted to increase by 0.43 points per year for each 1-point increase in a person's mean rapport rating (averaged across time points), holding all else constant. This | Relative to the average growth rate across both roles, novices' belongingness growth rate was predicted to increase by 0.22 points per year for each 1-point increase in their own average rapport rating, holding all else constant. This effect was not statistically significant ($p$ = .109). |

| | Fixed Effects | Dummy Coding Interpretation | Effect Coding Interpretation |
|---|---|---|---|
| | | three-way interaction was not statistically significant ($p = .109$). In other words, a person's typical rapport level did not significantly predict differences in growth rates between novices and experts. | That is, a person's trait-level rapport did not significantly moderate the role-based differences in growth trajectories. |
| 19 | Time*Role*Partner Rapport (WP) | The difference between novices' and experts' belongingness growth rates was predicted to increase by 0.05 points per year for each 1-point increase in a person's partner's rapport rating at that time point (relative to their partner's own average rapport), holding all else constant. This three-way interaction was not statistically significant ($p = .449$). In other words, momentary changes in a partner's rapport were not predictive of role-based differences in belongingness growth. | Relative to the average growth rate across both roles, novices' belongingness growth rate was predicted to decrease by 0.03 points per year for each 1-point increase in their partner's momentary rapport (relative to the partner's average), holding all else constant. This effect was not statistically significant ($p = .449$). That is, within-person fluctuations in a partner's rapport did not significantly moderate role differences in growth in belongingness. |
| 20 | Time*Role*Partner Rapport (Agg) | The difference between novices' and experts' belongingness growth rates was predicted to increase by 0.28 points per year for each 1-point increase in a person's partner's mean rapport rating (averaged across time points), holding all else constant. This three-way interaction was not statistically significant ($p = .293$). In other words, a partner's typical rapport level did not significantly predict role-based differences in belongingness growth over time. | Relative to the average growth rate across both roles, novices' belongingness growth rate was predicted to increase by 0.14 points per year for each 1-point increase in their partner's average rapport, holding all else constant. This effect was not statistically significant ($p = .130$). That is, differences in a partner's typical rapport level did not significantly moderate the relationship between role and growth in belongingness. |

| | Random Effects (Variances) | Dummy Coding Interpretation | Effect Coding Interpretation |
|---|---|---|---|
| 1 | Intercept variance | Dyad-level variability in belongingness at the midpoint of the study (Time = 0) for experts (role = 0) was estimated at Var = 16.52, indicating that approximately 15% of the variance in mid-study belongingness among experts was attributable to dyad membership. | Dyad-level variability in belongingness at the midpoint of the study, averaged across both roles, was estimated at Var = 9.23, indicating that approximately 24% of the variance in mid-study belongingness was attributable to dyad membership. |
| 2 | Slope (Time) variance | The variability in growth rates per year for experts (role = 0) was estimated at Var = 18.33, indicating that approximately 17% of the variance in belongingness was attributable to differences in experts' individual growth trajectories. | The variability in growth rates per year, averaged across both roles, was estimated at Var = 10.04, indicating that approximately 26% of the variance in belongingness was attributable to individual differences in change over time. |
| 3 | Role variance | The variance in belongingness attributable to role differences (novices vs. experts) at the midpoint of the study (Time = 0) was estimated at Var = 34.85, indicating that approximately 32% of the variance in belongingness was explained by systematic differences between roles within dyads. | The variance in belongingness attributable to role-related deviations from the grand mean at the midpoint of the study was estimated at Var = 8.71, indicating that approximately 22% of the variance in belongingness was explained by role differences within dyads. |
| 4 | Time*Role variance | The variability in growth rate differences between novices (role = 1) and experts (role = 0) was estimated at Var = 39.88, suggesting that approximately 36% of the variance in belongingness growth trajectories was attributable to role-based differences in change over time. | The variance in growth rate differences for novices (role = +1) relative to the average growth rate across roles was estimated at Var = 9.97, suggesting that approximately 26% of the variance in belongingness growth trajectories was attributable to role-based deviations from the average rate of change. |

Note. This table includes interpretations of fixed and random effects from Model 2 (hybrid APIM-CFGM), which incorporates both time-varying (within-person) and time-invariant (aggregated) actor and partner covariates. Dummy coding (expert = 0, novice = 1) was used in primary model estimation; effect coding (expert = +1, novice = -1) is included for interpretive comparison. All fixed effect interpretations are evaluated at the study midpoint (Time = 0), unless part of a time interaction. Random effect variances reflect variability across dyads, time slopes, or role contrasts, as specified. Within-person predictors were person-mean centered; aggregated predictors were grand-mean centered.

**Figure 1**

*Path Diagrams for Cross-Sectional Non-Exchangeable Dyadic Analyses*

**Panel A: Actor-Partner Interdependence Model (APIM)**          **Panel B: Common Fate Model (CFM)**

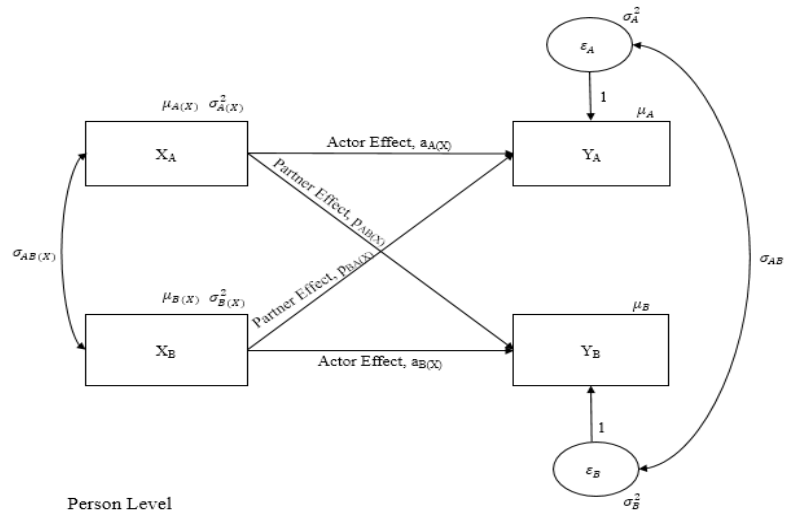

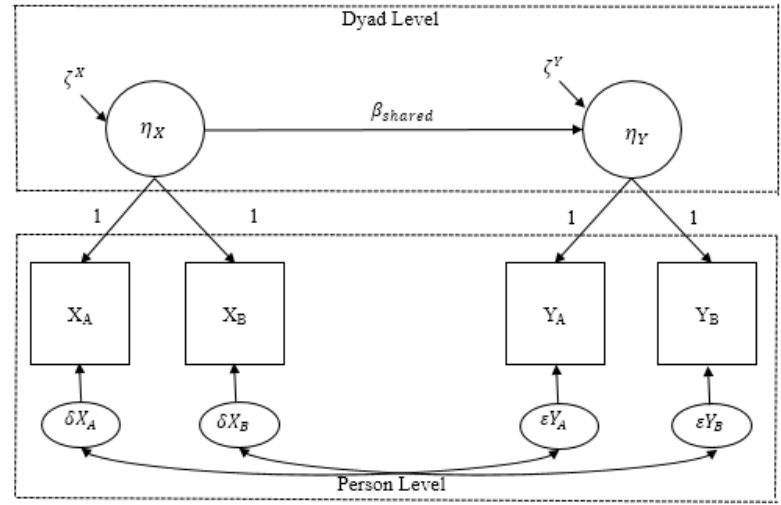

**Panel C: AP-CFM Hybrid**                                          **Panel D: CF-APM Hybrid**

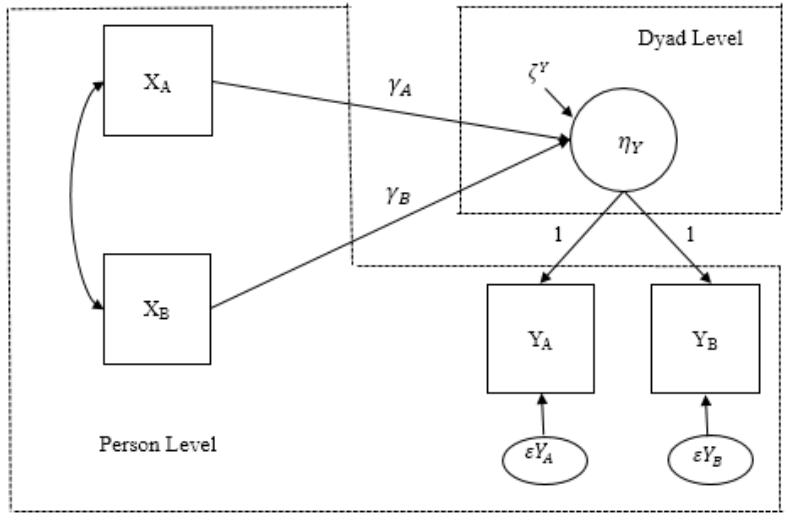

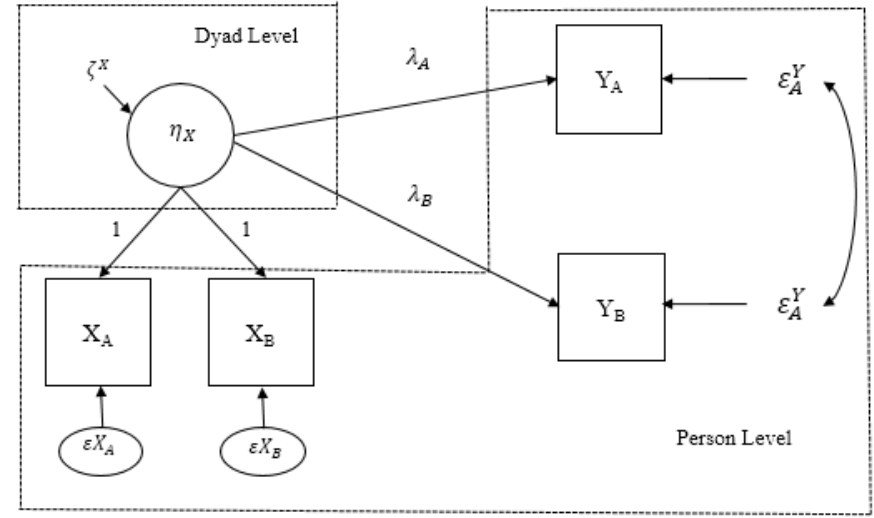

*Note.* All figures adapted from Wickham and Macia (2019). The two members of the dyad are represented by the index "A" and "B" for member A and member B in a dyad, respectively. Panel A and Panel B are for the $j_{th}$ dyad. Panel A - Diagrams for the APIM: $a_{A(X)}$ and $a_{B(X)}$ represent the effect of one's own characteristics on his/her outcome values; $p_{AB(X)}$ and $p_{BA(X)}$ represent the effect of one's partner characteristics on his/her own outcome values. The single-headed arrows indicate predictive relationships, and double-headed arrows represent dependencies. The double–headed arrow between ε represents the residual nonindependence in the outcome scores that can't be explained by APIM, thus by the predictors included in the model. Panel B - Diagrams for the CFM: the intercepts of indicator variable are freely estimated. The latent mean or intercept for the common fate variables ($\eta_X$ and $\eta_Y$) are fixed to 0. Panel C - Diagrams for the AP-CFM: The AP-CFM features a dyad-level latent outcome $\eta_Y$ regressed on each person-level predictor $X_A, X_B$. The latent outcome $\eta_Y$ is measured by both member's observed outcome $Y_A, Y_B$, capturing shared dyadic variance. Paths labeled γ regression coefficients from each person-level predictor to the dyad-level latent outcome. Subscripts indicate the source (actor or partner) and dyad member. Panel D - Diagrams for the CF-APM: The CF-APM features the individual-level outcome variables $Y_A, Y_B$ regressed on a dyad- level latent predictor $\eta_X$, which is measured by both members' observed predictor values $X_A, X_B$. This structure allows a common latent attribute to predict each partner's outcome. Paths labeled λ represents represent regression coefficients from the latent predictor to each person's outcome.

**Figure 2**

*Path Diagram of 5-Time Point Longitudinal Non-Exchangeable Dyadic Analyses using the Latent Common Fate Growth Model*

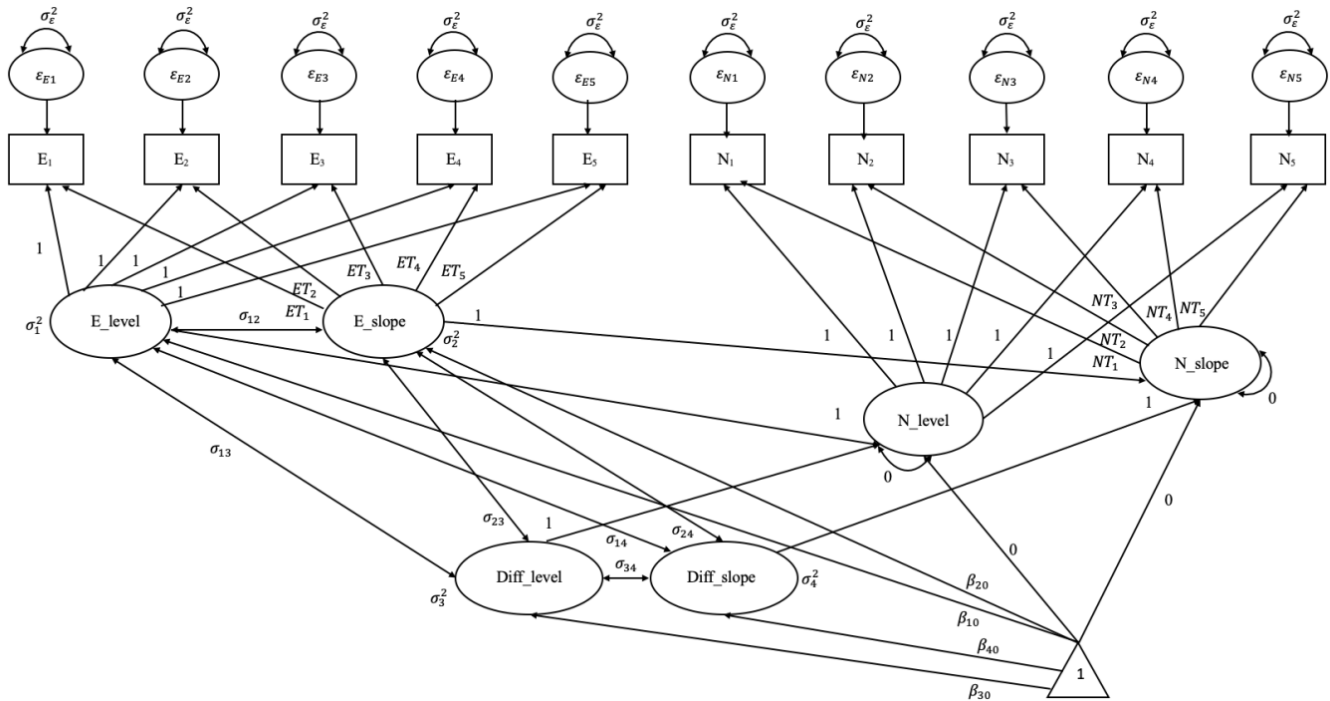

*Note.* A path diagram adapted from Planalp et al. (2017) of a linear trajectory Common Fate Growth Model (CFGM) with five occasions of measurement. This is the default error covariance structure. The two members of the dyad are represented by the index "E" and "N" for expert and novice in a dyad. The unique residual variances are constrained to be equal across the time points and between expert (E) and novice (N). The covariance between expert (E) and novice (N) unique residuals at each time point is constrained to 0. $T_1$, $T_2$, $T_3$, $T_4$, and $T_5$ are -0.75, -0.5, 0, 0.5, and 1 years.

**Figure 3**

*Data Structure for Hypothetical Example*

| dyadid | personid | belong | role_eff | role_dum | time | rapport_actor_ within | rapport_partner_ within | rapport_actor_ agg | rapport_partner_ agg |
|---|---|---|---|---|---|---|---|---|---|
| 1 | 1 | 1.11 | 1 | 1 | -0.75 | 1.65 | -3.38 | -0.30 | 4.41 |
| 1 | 1 | -0.98 | 1 | 1 | -0.50 | 0.27 | -2.55 | -0.30 | 4.41 |
| 1 | 1 | -1.40 | 1 | 1 | 0.00 | -0.37 | -0.36 | -0.30 | 4.41 |
| 1 | 1 | -1.68 | 1 | 1 | 0.50 | -0.67 | 2.87 | -0.30 | 4.41 |
| 1 | 1 | -3.11 | 1 | 1 | 1.00 | -0.88 | 3.43 | -0.30 | 4.41 |
| 1 | 51 | 1.28 | -1 | 0 | -0.75 | -3.38 | 1.65 | 4.41 | -0.30 |
| 1 | 51 | -0.16 | -1 | 0 | -0.50 | -2.55 | 0.27 | 4.41 | -0.30 |
| 1 | 51 | 1.77 | -1 | 0 | 0.00 | -0.36 | -0.37 | 4.41 | -0.30 |
| 1 | 51 | 3.56 | -1 | 0 | 0.50 | 2.87 | -0.67 | 4.41 | -0.30 |
| 1 | 51 | 1.67 | -1 | 0 | 1.00 | 3.43 | -0.88 | 4.41 | -0.30 |
| 2 | 2 | -2.16 | 1 | 1 | -0.75 | -0.31 | 1.18 | 4.04 | 2.59 |
| 2 | 2 | 1.35 | 1 | 1 | -0.50 | -1.83 | -1.27 | 4.04 | 2.59 |
| 2 | 2 | 3.57 | 1 | 1 | 0.00 | -0.86 | -0.97 | 4.04 | 2.59 |
| 2 | 2 | 5.84 | 1 | 1 | 0.50 | 0.77 | 0.00 | 4.04 | 2.59 |
| 2 | 2 | 7.74 | 1 | 1 | 1.00 | 2.22 | 1.06 | 4.04 | 2.59 |
| 2 | 52 | -0.64 | -1 | 0 | -0.75 | 1.18 | -0.31 | 2.59 | 4.04 |
| 2 | 52 | 1.42 | -1 | 0 | -0.50 | -1.27 | -1.83 | 2.59 | 4.04 |
| 2 | 52 | -0.03 | -1 | 0 | 0.00 | -0.97 | -0.86 | 2.59 | 4.04 |
| 2 | 52 | -0.50 | -1 | 0 | 0.50 | 0.00 | 0.77 | 2.59 | 4.04 |
| 2 | 52 | -0.14 | -1 | 0 | 1.00 | 1.06 | 2.22 | 2.59 | 4.04 |

**Figure 4**

*Parameter Estimates and Standard Errors with Larger vs. Smaller Samples of Dyads*

**Panel A: Coefficient Estimates**

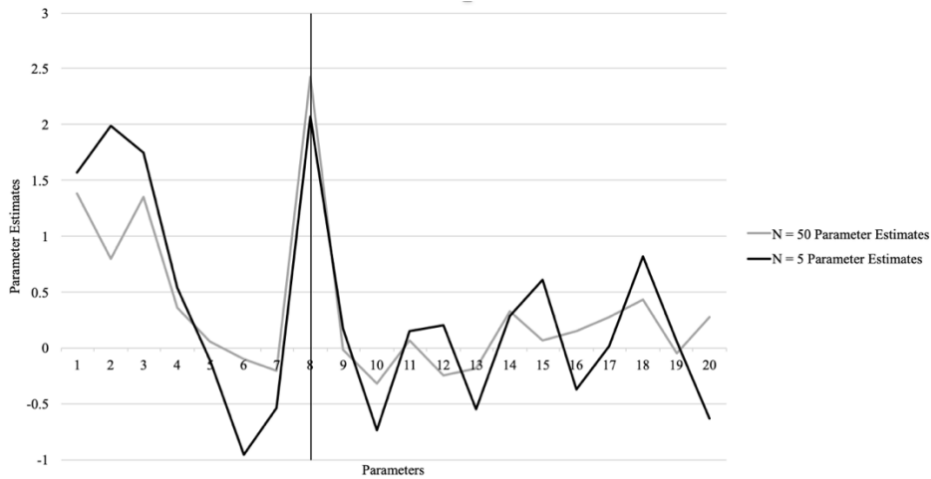

**Panel B: Coefficient Standard Error Estimates**

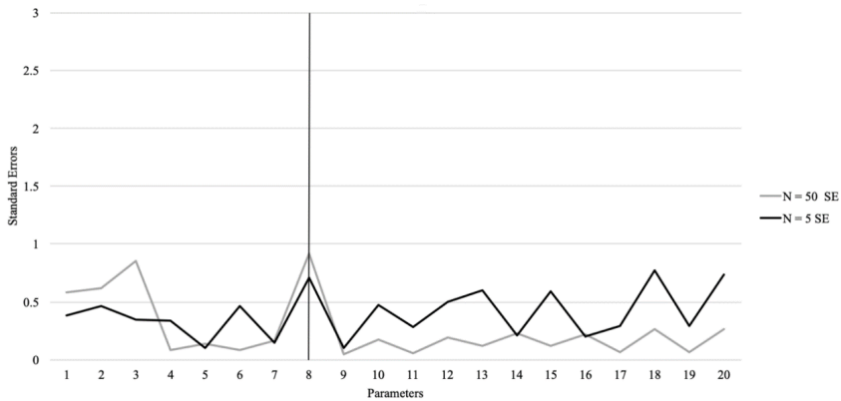

*Note.* Parameter estimates and standard errors by sample size (50 vs 5 dyads). Panel A displays selected fixed-effect estimates (coefficients) from the hybrid APIM-CFGM model (dummy-coded role) under ML estimation; Panel B shows the corresponding standard errors. The numbers along the horizontal axis represent parameter estimates corresponding to the numbers in Tables 1.

**Figure 5**

*Parameter Estimates and Standard Errors with Maximum Likelihood vs. Bayesian Estimation*

**Panel A: Coefficient Estimates**

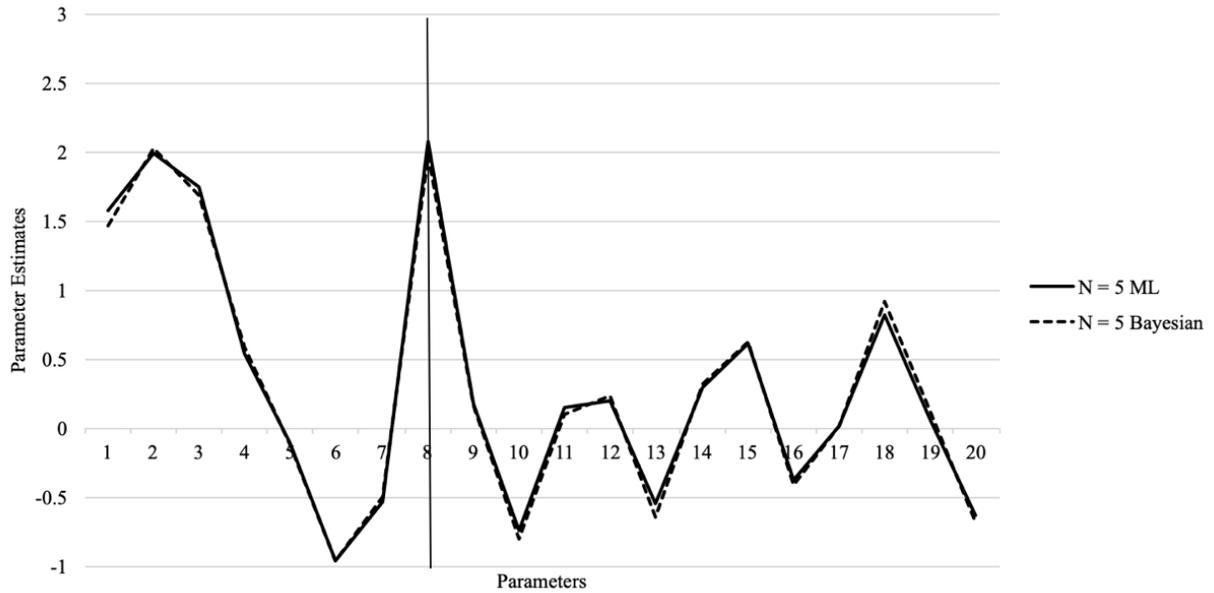

**Panel B: Standard Error Estimates**

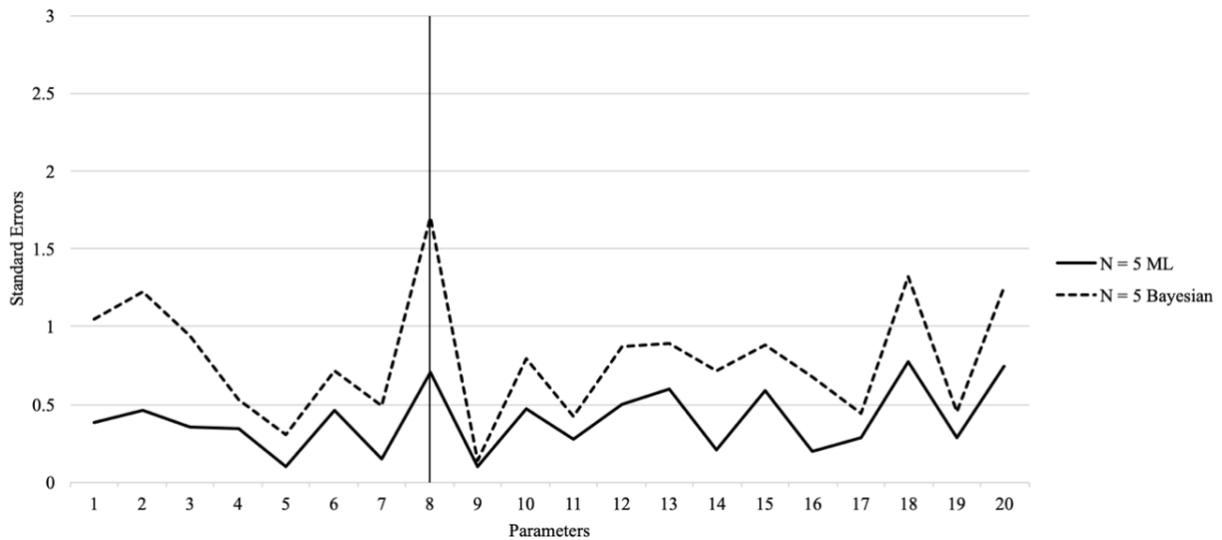

*Note.* Panel A plots ML coefficients against Bayesian posterior means for all fixed effects, with the diagonal line indicating perfect agreement. Panel B plots ML standard errors against Bayesian posterior standard deviations. The results are based on dummy-coded role. The numbers along the horizontal axis represent parameter estimates corresponding to the numbers in Tables 1.

**Appendix**

**Table A1**

*Interpretation of Model Parameters Under Dummy Coding vs. Effect Coding of the Role Variable*

| Term | Dummy Coding Interpretation | Effect Coding Interpretation |
|---|---|---|
| Intercept (Baseline outcome) | Mean outcome for the reference role coded 0 at baseline (Time = 0). | Mean outcome across both roles at baseline (Time = 0). |
| Time (Slope over time) | The predicted change in outcome over time for the reference role coded 0, holding all else constant. | Average change in outcome over time (slope) across both roles, holding all else constant. |
| Role | The difference in baseline outcome at Time = 0 between the role coded 1 and role coded 0, holding all else constant. | The deviation of each role's baseline outcome from the grand mean at Time = 0. The role coded +1 is above the mean by this amount; the role coded -1 has the same deviation in the opposite direction. |
| Actor Covariate (Within-Person) | At Time = 0, the predicted change in the outcome for the role coded 0 for each 1-unit increase in one's own covariate value, holding all else constant. | At Time = 0, the predicted change in the outcome across both roles for each 1-unit increase in one's own covariate value (deviation from their person mean), holding all else constant. |
| Actor Covariate (Aggregate) | At Time = 0, the predicted change in the outcome for the role coded 0 for each 1-unit increase in one's own mean covariate value across time points, holding all else constant. | At Time = 0, the predicted change in the outcome across both roles for each 1-unit increase in one's own mean covariate value across time points, holding all else constant. |
| Partner Covariate (Within-Person) | At Time = 0, the predicted change in the outcome for the role coded 0 for each 1-unit increase in one's partner's covariate deviation (relative to their person mean), holding all else constant. | At Time = 0, the predicted change in the outcome across both roles for each 1-unit increase in one's partner's covariate value (relative to their person mean), holding all else constant. |
| Partner Covariate (Aggregate) | At Time = 0, the predicted change in outcome for the role coded 0 for each 1-unit increase in one's partner's mean covariate value across time points, holding all else constant. | At Time = 0, the predicted change in the outcome across both roles for each 1-unit increase in one's partner's mean covariate value across time points, holding all else constant. |

| Term | Dummy Coding Interpretation | Effect Coding Interpretation |
|---|---|---|
| Time × Role | Difference in the rate of change (slope) over time between the role coded 1 and the reference role (coded 0), holding all else constant. | The difference in the rate of change (slope) over time for the role coded +1 compared to the average growth rate across both roles; the role coded -1 differs by the same amount in the opposite direction, holding all else constant. |
| Time × Actor Covariate (Within-Person) | Change in outcome slop over time for the reference role (coded 0) associated with a 1-unit increase in one's own covariate deviation (relative to their own person mean), holding all else constant. | The average change in the growth rate of the outcome over time across both roles, for each 1-unit increase in one's own covariate value (relative to their person mean), holding all else constant. |
| Time × Actor Covariate (Aggregate) | Change in the outcome growth rate over time for the role coded 0, for each 1-unit increase in one's own mean covariate value across time points, holding all else constant. | The average change in the outcome growth rate over time across both roles, for each 1-unit increase in one's own mean covariate value across time points, holding all else constant. |
| Time × Partner Covariate (Within-Person) | Change in the outcome growth rate over time for the role coded 0, for each 1-unit increase in one's partner's covariate deviation (relative to their person mean), holding all else constant. | The average change in the outcome growth rate over time across both roles, for each 1-unit increase in one's partner's covariate value (relative to their person mean), holding all else constant. |
| Time × Partner Covariate (Aggregate) | The change in the outcome growth rate over time for the role coded 0, for each 1-unit increase in one's partner's mean covariate value across time points, holding all else constant. | The average change in the outcome growth rate over time across both roles, for each 1-unit increase in one's partner's mean covariate value across time points, holding all else constant. |
| Role × Actor Covariate (Within-Person) | The difference between role coded 1 and the reference role (coded 0) in the effect of one's own time-varying covariate (relative to their person mean) on the outcome at Time = 0, holding all else constant. | The difference between role coded +1 and the average relationship across both roles in the effect of one's own time-varying covariate (relative to their person mean) on the outcome at Time = 0; the role coded -1 differs by the same amount in the opposite direction, holding all else constant. |
| Role × Actor Covariate (Aggregate) | The difference between role coded 1 and the reference role (coded 0) in the effect of one's own mean covariate value across time points on | The difference between role coded +1 and the average effect across both roles in the relationship between one's own mean covariate value and the |

| Term | Dummy Coding Interpretation | Effect Coding Interpretation |
|---|---|---|
| | the outcome at Time = 0, holding all else constant. | outcome at Time = 0; the role coded -1 differs by the same amount in the opposite direction, holding all else constant. |
| Role × Partner Covariate (Within-Person) | The difference between role coded 1 and the reference role (coded 0) in the effect of one's partner's time-varying covariate (relative to their person mean) on the outcome at Time = 0, holding all else constant. | The difference between role coded +1 and the average effect across both roles in the relationship between one's partner's time-varying covariate (relative to their person mean) and the outcome at Time = 0; the role coded -1 differs by the same amount in the opposite direction, holding all else constant. |
| Role × Partner Covariate (Aggregate) | The difference between role coded 1 and the reference role (coded 0) in the effect of one's partner's mean covariate value across time points on the outcome at Time = 0, holding all else constant. | The difference between role coded +1 and the average effect across both roles in the relationship between one's partner's mean covariate value and the outcome at Time = 0; the role coded -1 differs by the same amount in the opposite direction, holding all else constant. |
| Time × Role × Actor Covariate (Within-Person) | The difference between role coded 1 and the reference role (coded 0) in how one's own time-varying covariate (relative to their person mean) predicts change in the outcome over time, holding all else constant. | The difference between role coded +1 and the average relationship across both roles in how one's own time-varying covariate (relative to their person mean) predicts change in the outcome over time; the role coded -1 differs by the same amount in the opposite direction, holding all else constant. |
| Time × Role × Actor Covariate (Aggregate) | The difference between role coded 1 and the reference role (coded 0) in how one's own mean covariate value across time points predicts change in the outcome over time, holding all else constant. | The difference between role coded +1 and the average relationship across both roles in how one's own mean covariate value predicts change in the outcome over time; the role coded -1 differs by the same amount in the opposite direction, holding all else constant. |
| Time × Role × Partner Covariate (Within-Person) | The difference between role coded 1 and the reference role (coded 0) in how one's partner's time-varying covariate (relative to their person | The difference between role coded +1 and the average relationship across both roles in how one's partner's time-varying covariate (relative to their person mean) predicts change in the outcome over |

| Term | Dummy Coding Interpretation | Effect Coding Interpretation |
|---|---|---|
| | mean) predicts change in the outcome over time, holding all else constant | time; the role coded -1 differs by the same amount in the opposite direction, holding all else constant. |
| Time × Role × Partner Covariate (Aggregate) | The difference between role coded 1 and the reference role (coded 0) in how one's partner's mean covariate value across time points predicts change in the outcome over time, holding all else constant. | The difference between role coded +1 and the average relationship across both roles in how one's partner's mean covariate value predicts change in the outcome over time; the role coded -1 differs by the same amount in the opposite direction, holding all else constant. |

*Note.* Dummy coding compares role = 1 to a reference group (role = 0). Effect coding uses role = +1 vs. -1, so each effect under effect coding is interpreted as a deviation from the grand mean rather than from a reference group. Terms are described at baseline (*Time* = 0) for intercepts and covariate effects, and as changes in slopes for interactions with Time. Higher-order interactions (Time × Role × Covariate) represent moderation of effects by both role and time.